\documentclass[]{mn2e}
\usepackage{epsfig,lscape}
\def\gsim{\vcenter{\hbox{$>$}\offinterlineskip\hbox{$\sim$}}}
\def\lsim{\vcenter{\hbox{$<$}\offinterlineskip\hbox{$\sim$}}}
\title[ISM in front of $\omega$\,Cen]{Detailed maps of interstellar clouds in
front of $\omega$\,Centauri: Small-scale structures in the Galactic Disc--Halo
interface}
\author[van Loon et al.]{Jacco Th. van Loon$^{1}$, Keith T.\ Smith$^{2}$, Iain
McDonald$^{1}$, Peter J.\ Sarre$^{2}$,
\newauthor Stephen J.\ Fossey$^{3}$ and Robert G.\ Sharp$^{4}$\\
$^{1}$Astrophysics Group, Lennard-Jones Laboratories, Keele University,
Staffordshire ST5 5BG, United Kingdom\\
$^{2}$School of Chemistry, The University of Nottingham, Nottingham NG7 2RD,
United Kingdom\\
$^{3}$Department of Physics and Astronomy, University College London, London
WC1E 6BT, United Kingdom\\
$^{4}$Anglo-Australian Observatory, P.O.\ Box 296, Epping, NSW 1710, Australia}
\date{Submitted on 3 April 2009}
\pagerange{\pageref{firstpage}--\pageref{lastpage}}
\pubyear{2009}
\begin{document}
\maketitle
\label{firstpage}
\begin{abstract}
The multi-phase interstellar medium (ISM) is highly structured, on scales from
the size of the Solar System to that of a Galaxy. In particular small-scale
structures are difficult to study and hence are poorly understood. We
used the multiplex capabilities of the AAOmega spectrograph at the
Anglo-Australian Telescope to create a half-square-degree map of the neutral
and low-ionized ISM in front of the nearby ($\sim5$ kpc), most massive
Galactic globular cluster, $\omega$\,Centauri. Its redshifted, metal-poor and
hot horizontal branch stars probe the medium-strong Ca\,{\sc ii} K and
Na\,{\sc i} D$_2$ line absorption, and weak absorption in the $\lambda$5780
and $\lambda$5797 Diffuse Interstellar Bands (DIBs), on scales around a
parsec. The kinematical and thermodynamical picture emerging from these data
is that we predominantly probe the warm neutral medium and weakly-ionized
medium of the Galactic Disc--Halo interface, $\sim0.3$--1 kpc above the
mid-plane. A comparison with {\it Spitzer Space Telescope} 24-$\mu$m and
DIRBE/IRAS maps of the warm and cold dust emission confirms that both Na\,{\sc
i} and Ca\,{\sc ii} trace the overall column density of the warm neutral and
weakly-ionized medium. Clear signatures are seen of the depletion of calcium
atoms from the gas phase into dust grains. Curiously, the coarse DIRBE/IRAS
map is a more reliable representation of the relative reddening between
sightlines than the Na\,{\sc i} and Ca\,{\sc ii} absorption-line measurements,
most likely because the latter are sensitive to fluctuations in the local
ionization conditions. The behaviour of the DIBs is consistent with the
$\lambda$5780 band being stronger than the $\lambda$5797 band in regions where
the ultraviolet radiation level is relatively high, as in the Disc--Halo
interface. This region corresponds to a $\sigma$-type cloud in which Ca\,{\sc
i} and small diatomic molecules such as CH and CN are usually absent. In all,
our maps and simple analytical model calculations show in unprecedented detail
that small-scale density and/or ionization structures exist in the
extra-planar gas of a spiral galaxy.
\end{abstract}

\section{Introduction}

``Overhead the sky was full of strings and shreds of vapour, flying,
vanishing, reappearing, and turning about an axis like tumblers, as the wind
hounded them through heaven.'' (Stevenson 1879). The interstellar medium (ISM)
is indeed an extra-ordinarily dynamic place. Cooling and heating processes are
continuously at work to transform gas between different states, from hot
ionized gas with $T\sim10^6$ K visible in X-rays, to a cold neutral medium
with $T<100$ K visible in H\,{\sc i} and far-IR emission (Heiles \& Troland
2005). The hotter media tend to fill space, whereas the colder media tend to
form discrete structures (Cox 2005). The dynamics are important for Galactic
recycling such as the re-accretion of gas from the Galactic Halo as fountains
of hot gas from supernovae in the Galactic Disc cool and condense. They also
influence the formation of small-scale structure in the form of filaments and
small clumps, which must be transient features as they are out of pressure
equilibrium with their surroundings (Stanimirovi\'c \& Heiles 2005). Some of
these structures have AU scales; we know very little about the 3D shapes of
these tiny-scale structures, although a sheet-like geometry seems likely
(Heiles \& Troland 2005).

Absorption-line studies, e.g., against pulsars in radio or stars in
optical/ultraviolet, can probe the small-scale structure; by using the
relative proper motion of the background source, 1D tiny-scale structure can
be traced (Heiles 1997; Crawford 2003). Mapping small- and tiny-scale
structure directly is often limited by large separations between background
sources, and limited angular resolution. Together with mapping high-surface
brightness nebulae (e.g., Lazio et al.\ 2009), globular clusters offer some of
the best possibilities for mapping experiments.

Langer, Prosser \& Sneden (1990) observed several sightlines towards the
globular clusters M\,15 and M\,92, finding evidence for significant column
density and velocity variations on scales as small as 0.2 pc. Bates et al.\
(1995) presented similar findings for sightlines towards the globular cluster
M\,13. Variations on even smaller scales, down to $\sim0.01$ pc, were detected
towards globular clusters M\,4 (Kemp, Bates \& Lyons 1993) and the central
parts of M\,15 (Meyer \& Lauroesch 1999) and M\,92 (Andrews, Meyer \&
Lauroesch 2001). The latter team points out that fluctuations in ionization
equilibrium and not total column density may be responsible for the observed
variations. Points, Lauroesch \& Meyer (2004) observed 150 sightlines towards
the binary open cluster h \& $\chi$\,Persei, detecting both small-scale
fluctuations as well as correlated behaviour over larger scales, suggesting
variations in the physical conditions {\it within} sheets of gas.

We are incredibly fortunate to find ourselves in close proximity to
$\omega$\,Centauri, the most massive Galactic globular cluster
($M\sim2.5\times10^6$ M$_\odot$; van de Ven et al.\ 2006). Few globular
clusters are nearer to the Sun than $\omega$\,Cen ($d=5$ kpc; McDonald et al.\
2009), resulting in a large number of suitable, bright probes of angular
scales between an arcsecond and a degree --- (sub)parsec scales in the
intervening ISM. We here use its many horizontal branch (HB) stars. These
stars are ideal probes: they are metal-poor, $-1.8<$[Fe/H]$<-0.8$, and hot,
$T>7000$ K, and thus relatively clear of photospheric absorption in the
interstellar lines of low excitation that probe the neutral and low-ionized
ISM. The high retrograde velocity of $\omega$\,Cen ($v_{\rm LSR}\sim229$
km~s$^{-1}$) moves the photospheric absorption out of the interstellar column.
The proper motions of the stars are known too ($\sim5$ mas~yr$^{-1}$, van
Leeuwen et al.\ 2000), so repeat observations may be used to probe the
tiny-scale structure.

%
%
\begin{table*}
\caption[]{Catalogue of observed stars in $\omega$\,Cen. Only the first eight
rows are displayed; the full catalogue of 452 objects is made available at
CDS. The first columns list the stellar probe's LEID number (van Leeuwen et
al.\ 2000) and its equatorial and Galactic coordinates, and the reddening
E(B--V) from the DIRBE/IRAS maps of Schlegel et al.\ (1998). Subsequent
columns list the central velocity with respect to the Local Standard of Rest,
line width ($\sigma$) and equivalent width ($W$) of the Ca\,{\sc ii} K and
Na\,{\sc i} D$_2$ interstellar lines, and the equivalent widths of the
$\lambda$5780 and $\lambda$5797 Diffuse Interstellar Bands.}
{\small
\begin{tabular}{lccccccccccccc}
\hline\hline
Star                                                              &
\multicolumn{2}{c}{\llap{E}quatorial coordinates}                 &
\multicolumn{2}{c}{\llap{G}alactic coordinate\rlap{s}}            &
\llap{E}(B--V\rlap{)}                                             &
\multicolumn{3}{c}{\llap{---}------ Ca\,{\sc i} K -----------}    &
\multicolumn{3}{c}{\llap{---}----- Na\,{\sc i} D$_2$ -----------} &
$\lambda$5780                                                     &
$\lambda$5797                                                     \\
                 &
$\alpha$ (J2000) &
$\delta$ (J2000) &
$l$              &
$b$              &
                 &
$v$              &
$\sigma$         &
$W$              &
$v$              &
$\sigma$         &
$W$              &
$W$              &
$W$              \\
LEID                 &
(deg)                &
(deg)                &
(deg)                &
(deg)                &
(mag)                &
\llap{(k}m/s\rlap{)} &
\llap{(}m\AA\rlap{)} &
(m\AA)               &
\llap{(k}m/s\rlap{)} &
\llap{(}m\AA\rlap{)} &
(m\AA)               &
(m\AA)               &
(m\AA)               \\
\hline
40004              &
201.06673          &
\llap{$-$}47.45981 &
308.669            &
\llap{1}5.0463     &
0.1301             &
\llap{$-$}19.4     &
28.6               &
\llap{5}$25\pm16$  &
\llap{$-$}13.6     &
22.7               &
\llap{4}$74\pm15$  &
$82\pm15$          &
---                \\
42009              &
201.11112          &
\llap{$-$}47.47200 &
308.698            &
\llap{1}5.0302     &
0.1304             &
\llap{$-$}32.1     &
30.3               &
\llap{5}$20\pm18$  &
\llap{$-$}13.2     &
23.0               &
\llap{4}$18\pm12$  &
$59\pm24$          &
---                \\
31006              &
201.12876          &
\llap{$-$}47.39312 &
308.721            &
\llap{1}5.1068     &
0.1256             &
\llap{$-$}24.3     &
29.6               &
\llap{5}$21\pm15$  &
\llap{$-$}13.4     &
23.6               &
\llap{3}$75\pm12$  &
$75\pm16$          &
---                \\
64010              &
201.18389          &
\llap{$-$}47.63934 &
308.726            &
\llap{1}4.8578     &
0.1417             &
\llap{$-$}27.9     &
32.2               &
\llap{6}$27\pm54$  &
\llap{$-$}16.5     &
26.2               &
\llap{5}$38\pm50$  &
$52\pm22$          &
\llap{3}$7\pm19$   \\
20006              &
201.14240          &
\llap{$-$}47.30854 &
308.743            &
\llap{1}5.1894     &
0.1227             &
\llap{$-$}17.8     &
27.2               &
\llap{4}$94\pm16$  &
\llap{$-$}12.1     &
22.6               &
\llap{3}$93\pm13$  &
$95\pm25$          &
\llap{4}$3\pm15$   \\
38008              &
201.18022          &
\llap{$-$}47.44549 &
308.750            &
\llap{1}5.0503     &
0.1286             &
\llap{$-$}25.4     &
29.9               &
\llap{5}$86\pm18$  &
\llap{$-$}17.3     &
27.0               &
\llap{5}$39\pm11$  &
---                &
---                \\
26009              &
201.19651          &
\llap{$-$}47.35413 &
308.774            &
\llap{1}5.1393     &
0.1227             &
\llap{$-$}23.9     &
27.7               &
\llap{5}$17\pm15$  &
\llap{$-$}14.0     &
25.3               &
\llap{3}$51\pm11$  &
$57\pm17$          &
\llap{2}$5\pm10$   \\
35011              &
201.21149          &
\llap{$-$}47.42335 &
308.775            &
\llap{1}5.0694     &
0.1272             &
\llap{$-$}20.6     &
29.7               &
\llap{5}$75\pm15$  &
\llap{$-$}18.3     &
25.4               &
\llap{5}$59\pm17$  &
$82\pm19$          &
---                \\
...                &
...                &
...                &
...                &
...                &
...                &
...                &
...                &
...                &
...                &
...                &
...                &
...                &
...                \\
\hline
\end{tabular}}
\end{table*}

The modest extinction, $E(B-V)\sim0.08$ mag (McDonald et al.\ 2009), and
$15^\circ$ Galactic latitude of $\omega$\,Cen make the interstellar path
towards $\omega$\,Cen relatively simple. Yet several components have been
identified, from the Local Bubble to a more distant spiral arm (Wood \& Bates
1994). The interesting possibility is that some of the material we would probe
could be in the Disc--Halo interface or associated with $\omega$\,Cen itself;
emission is seen at velocities of $\sim195$ km~s$^{-1}$ in H\,{\sc i}
(McDonald 2009; McDonald et al., in prep.) and CO (Origlia et al.\ 1997).
Absorption from the ISM in the 3968.47 \& 3933.68 \AA\ Ca\,{\sc ii} H \& K
doublet was easily detected in 2dF spectra (van Loon et al.\
2007)\footnote{Jones (1968) already noted the appearance of interstellar
K-line absorption in the spectrum of the F-type giant $\omega$\,Cen\,V1.}. The
same would be expected for the 5895.92 \& 5889.95 Na\,{\sc i} D$_1$ \& D$_2$
doublet. The 2dF spectra were of too low a resolution to resolve the
interstellar lines. We have since used its successor, AAOmega, at $R\sim8000$,
to measure ISM features in 452 hot HB stars in $\omega$\,Cen, of which we
present the results here.

\section{Observations}

\subsection{Spectroscopy with AAOmega}

The observations were undertaken, in service mode, at the Anglo-Australian
Telescope (AAT) on the night of 28 March 2008. The AAOmega spectrograph was
used with the multiple-object-spectroscopy fibre feed from the 2dF fibre
positioner (Saunders et al.\ 2004; Sharp et al.\ 2006).

The 5700 \AA\ AAOmega dichroic was used. The red and blue arms of AAOmega were
used with the 3200B and 2000R VPH gratings centered and blazed at 4040 \AA\
and 5990 \AA\ yielding spectra in the 3890--4170 \AA\ and 5740--6220 \AA\
regions at spectral resolutions (per 3.4-pixel resolution element) of
$R\sim8200$ ($\Delta v\sim37$ km s$^{-1}$) and $R\sim7245$ ($\Delta v\sim41$
km s$^{-1}$), respectively. The exact ranges of individual spectra differ. The
Ca\,{\sc ii} H \& K lines, which were the primary targets for the blue arm
observations, were placed off centre on the CCD to avoid regions of poor
cosmetic performance (bad columns). Skies were clear with seeing varying
during the night from $1.4^{\prime\prime}$ to $2.0^{\prime\prime}$; the 2dF
fibre entrances subtend $2.0^{\prime\prime}$ on the sky.

Two fibre configurations were employed. Each observing block consists of a
flat field frame (quartz-halogen lamp), an arc frame for wavelength
calibration (CuAr, FeAr, He and Ne), a set of twilight sky flat field frames
(to normalize the relative fibre transmissions for sky subtraction) and a
series of 1800 second science frames. The total observing times for the two
configurations sampling different stars were $6\times1800$ seconds and
$4\times1800+1\times1200$ seconds, respectively.

The data were processed using the {\sc 2dfdr} data reduction package. Two
additional steps were performed under {\sc iraf} prior to processing with {\sc
2dfdr}. Firstly, a 2D bias frame was subtracted to correct a number of poor
but recoverable columns in the data set. The 2D bias frame was created for
this purpose from a stack of 30 bias frames {\it via} a $\sigma$-clipped mean
rejection. Secondly, the {\sc iraf fixpix} task was used to interpolate across
a single bad column at the wavelength of the Ca\,{\sc ii} K line. This
isolated bad column affected the Ca\,{\sc ii} K line in around 25\% of
spectra, with spectral curvature across the CCD moving it clear for the
remaining stars. An updated bad-pixel mask was also created for {\sc 2dfdr} to
account for the effects of these modifications.

{\sc 2dfdr} performs the standard reduction operations: overscan correction;
fibre trace and extraction; flat fielding and wavelength calibration. Cosmic
ray rejection was applied to the 2D science frames by {\sc 2dfdr} prior to
extraction, following the prescription of van Dokkum (2001). In each
configuration, 25 fibres were devoted to blank sky regions to create an
average spectrum of the sky. Sky subtraction was performed using this sky
spectrum. The relative intensity of the sky to subtract was determined from
twilight flat-field frames obtained at the end of the night. Science frames
were combined using a relative flux weighting derived from spectral
intensities in each frame.

The spectra were not Doppler-corrected, i.e.\ they reflect the observer's
motion through space. All velocities used in this work and in the catalogue
have been corrected to the Local Standard of Rest (LSR), though, by
subtracting the $-10$ km~s$^{-1}$ motion of the observer with respect to the
LSR at the time of observation.

The continuum signal-to-noise ratio as measured on the extracted spectra is
typically 30--50 per pixel around the Ca\,{\sc ii} K line and 60--100 per
pixel around the Na\,{\sc i} D$_2$ line.

\subsection{Spectroscopic targets}

Targets were selected from the proper motion catalogue of van Leeuwen et al.\
(2000). The criteria were designed to select blue stars, with $14.6<B<16.6$
mag and $-0.2<(B-V)<0.25$ mag, that are likely members of $\omega$\,Cen, with
a membership probability $p\geq90$\% (all stars actually observed had
$p\geq95$\%, and mostly 99 or 100\%). A small number of targets were rejected
due to close neighbours ($<3^{\prime\prime}$), which would lead to
contaminating light within the $2^{\prime\prime}$-diameter fibres. Hence 995
potential targets were available.

To maximise the uniformity and density of the area on the sky covered by our
observations, targets were drawn from a field with radius $R=0.5^\circ$ and
mutual separation $>1^\prime$. The bluest stars, with $-0.20<(B-V)<0.10$ mag,
were given priority, then stars with $0.10\leq(B-V)<0.15$ mag at
$R>0.15^\circ$, followed by stars with $0.15\leq(B-V)<0.20$ mag at
$R>0.20^\circ$ and stars with $0.20\leq(B-V)\leq0.25$ mag at $R>0.25^\circ$.
We used the {\sc configure} software with the simulated annealing method to
configure the fields, and hence produced a second field with the leftovers
from the first configuration. In all, 452 targets were observed in two field
configurations.

To position the fibres accurately on sky, 6 and 7 fiducial stars were observed
for the first and second configuration, respectively. These were chosen from
amongst the brighter red giants, with $12.8<V<13.3$ mag and $0.9<(B-V)<1.3$
mag, with the same membership criterion as before to guarantee a good relative
astrometric and fibre-positioning accuracy. The sky positions were chosen
using the same target selection software, avoiding contamination within
$3^{\prime\prime}$ equivalent to $V=21$ mag in the fibre.

The observed stars are listed in Table 1, the full version of which is made
available electronically; their locations in the B, B--V colour--magnitude
diagram and on the sky are shown in Figs.\ 1 and 2, respectively.

%
%
\begin{figure}
\centerline{\psfig{figure=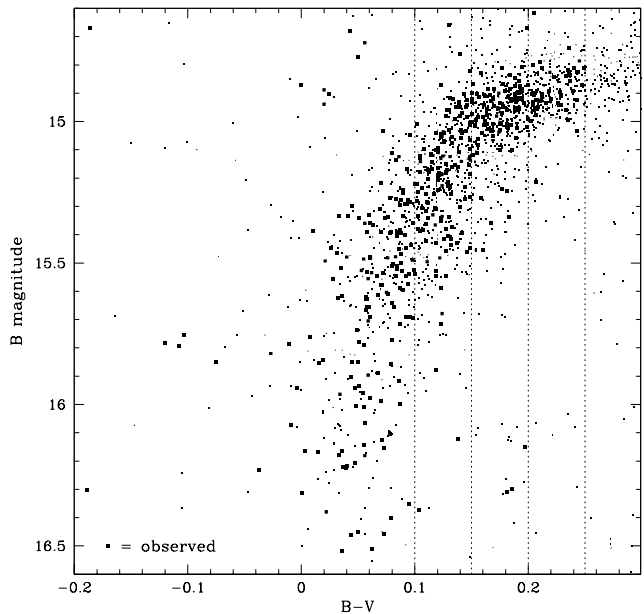,width=84mm}}
\caption[]{Optical B, B--V colour--magnitude diagram of potential targets
(small dots) with $\geq50$\% membership probability, and observed stars
(squares). The vertical dotted lines indicate priorities, with bluer stars
given higher priority.}
\end{figure}

%
%
\begin{figure}
\centerline{\psfig{figure=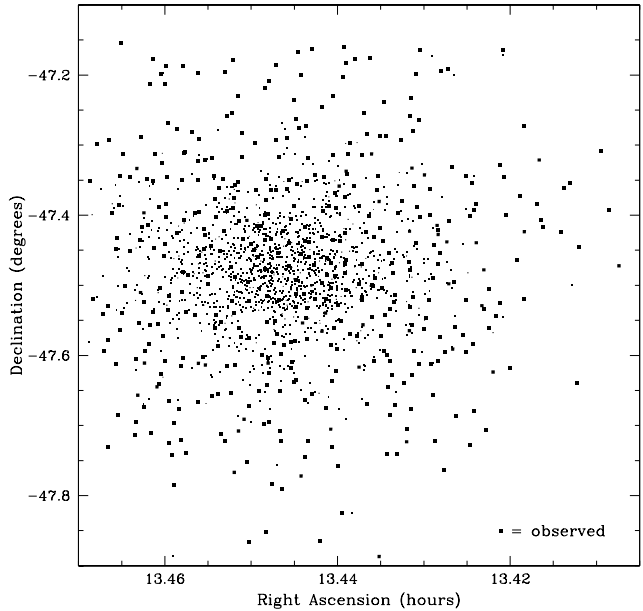,width=84mm}}
\caption[]{Positional diagram of potential targets (small dots) with
$\geq50$\% membership probability, $14.6<B<16.6$ mag and $-0.2<(B-V)<0.25$
mag, and observed stars (squares).}
\end{figure}

\section{Results}

%
%
\begin{figure}
\centerline{\psfig{figure=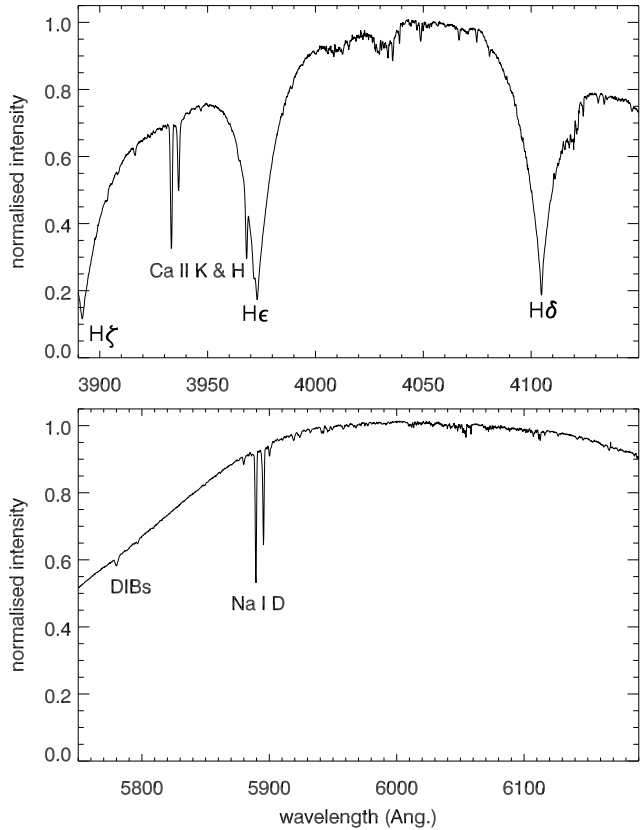,width=84mm}}
\caption[]{Overview of the average spectrum, in the blue ({\it top}) and red
({\it bottom}) regions, uncorrected for the spectral throughput and response
of the atmosphere, telescope and instrument. Note that the Ca\,{\sc ii} H
absorption features overlap the strong photospheric H$\epsilon$ line.}
\end{figure}

The average spectra in the blue and red regions are shown in Fig.\ 3. Besides
the strong Balmer hydrogen absorption lines arising in the photospheres of the
stars, the interstellar Ca\,{\sc ii} H \& K and Na\,{\sc i} D$_1$ \& D$_2$
lines are very obvious features observed with ease in every single spectrum.
Two Diffuse Interstellar Bands (DIBs) are discernible in the red portion of
the average spectrum, at 5780 and 5797 \AA.

\subsection{Interstellar Ca\,{\sc ii} and Na\,{\sc i} absorption}

Average profiles of the Ca\,{\sc ii} H \& K and Na\,{\sc i} D$_1$ \& D$_2$
lines are shown in Fig.\ 4: the shorter-wavelength interstellar absorption
lines of these multiplets, Ca\,{\sc ii} K and Na\,{\sc i} D$_2$, are well
isolated from their photospheric counterparts and are seen against a clean
photospheric continuum; the weaker Ca\,{\sc ii} H and Na\,{\sc i} D$_1$ lines
are affected by strong photospheric Balmer H$\epsilon$ and photospheric
Na\,{\sc i} D$_2$ line absorption, respectively. The kinematic offset between
stars and ISM is small enough to avoid confusion with the 5875.6 \AA\ He\,{\sc
i} D$_3$ line.

%
%
\begin{figure}
\centerline{\vbox{
\psfig{figure=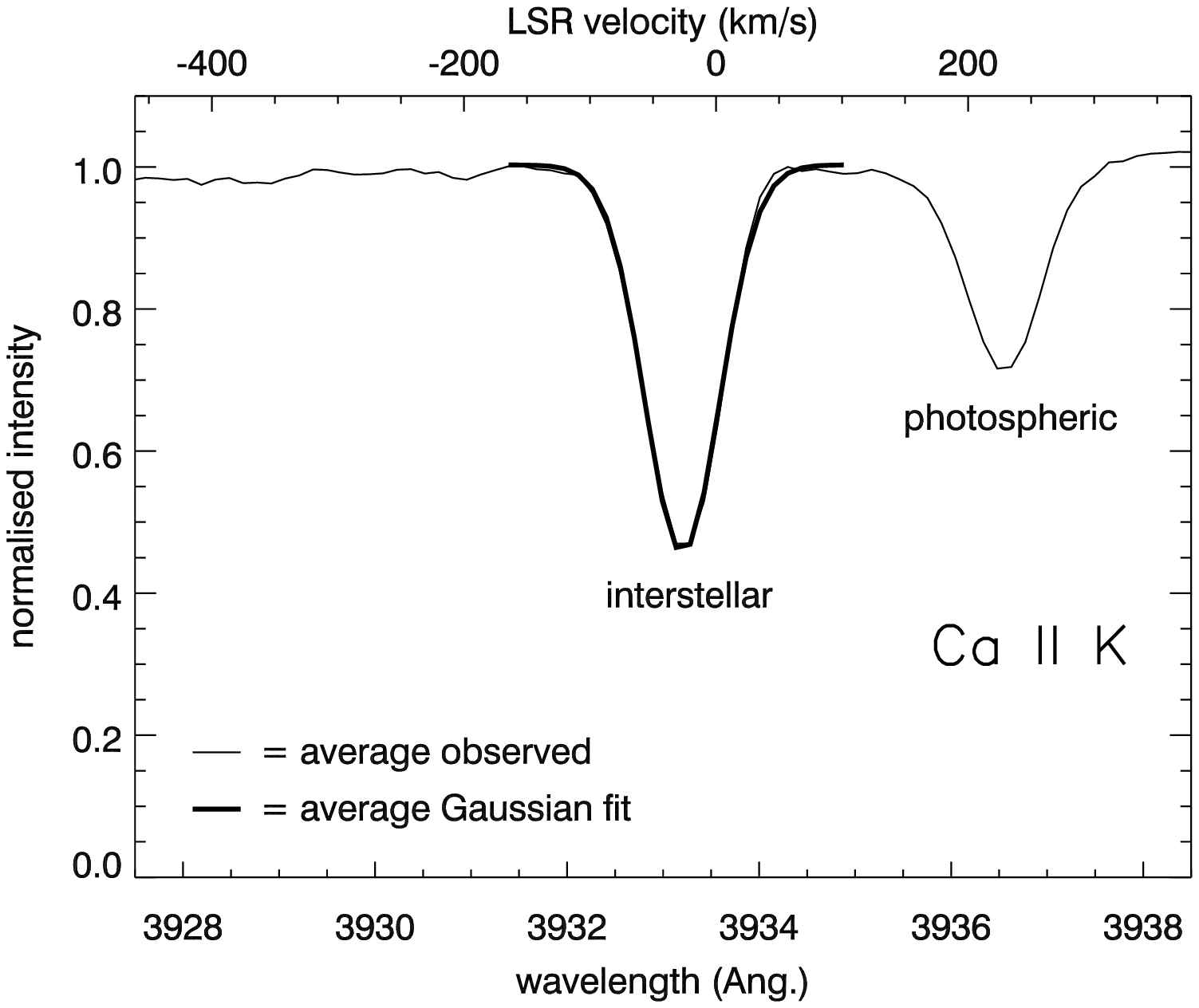,width=84mm}
\psfig{figure=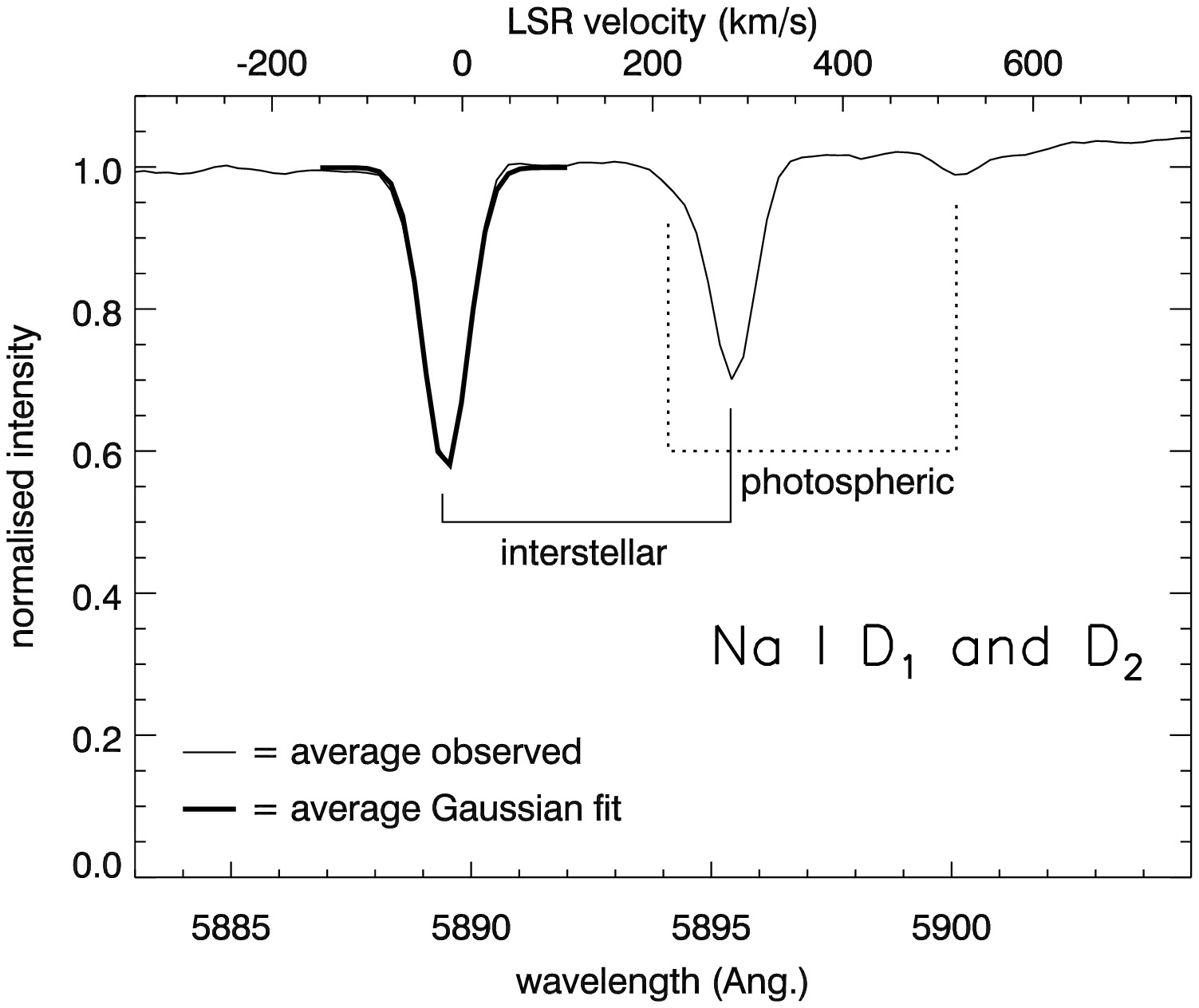,width=84mm}
}}
\caption[]{{\it Top}: The average spectrum around the Ca\,{\sc ii} K line. The
interstellar feature is well separated from the redshifted photospheric
absorption. The average Gaussian fit is overlaid, demonstrating the excellence
of the fits. {\it Bottom}: The average spectrum around the Na\,{\sc i} D$_1$
and D$_2$ lines. The interstellar D$_2$ line is well isolated, but the D$_1$
line is affected by the (generally much weaker) photospheric D$_2$ absorption.
The interstellar line profiles are very well matched by single Gaussians.}
\end{figure}

%
%
\begin{figure*}
\centerline{\psfig{figure=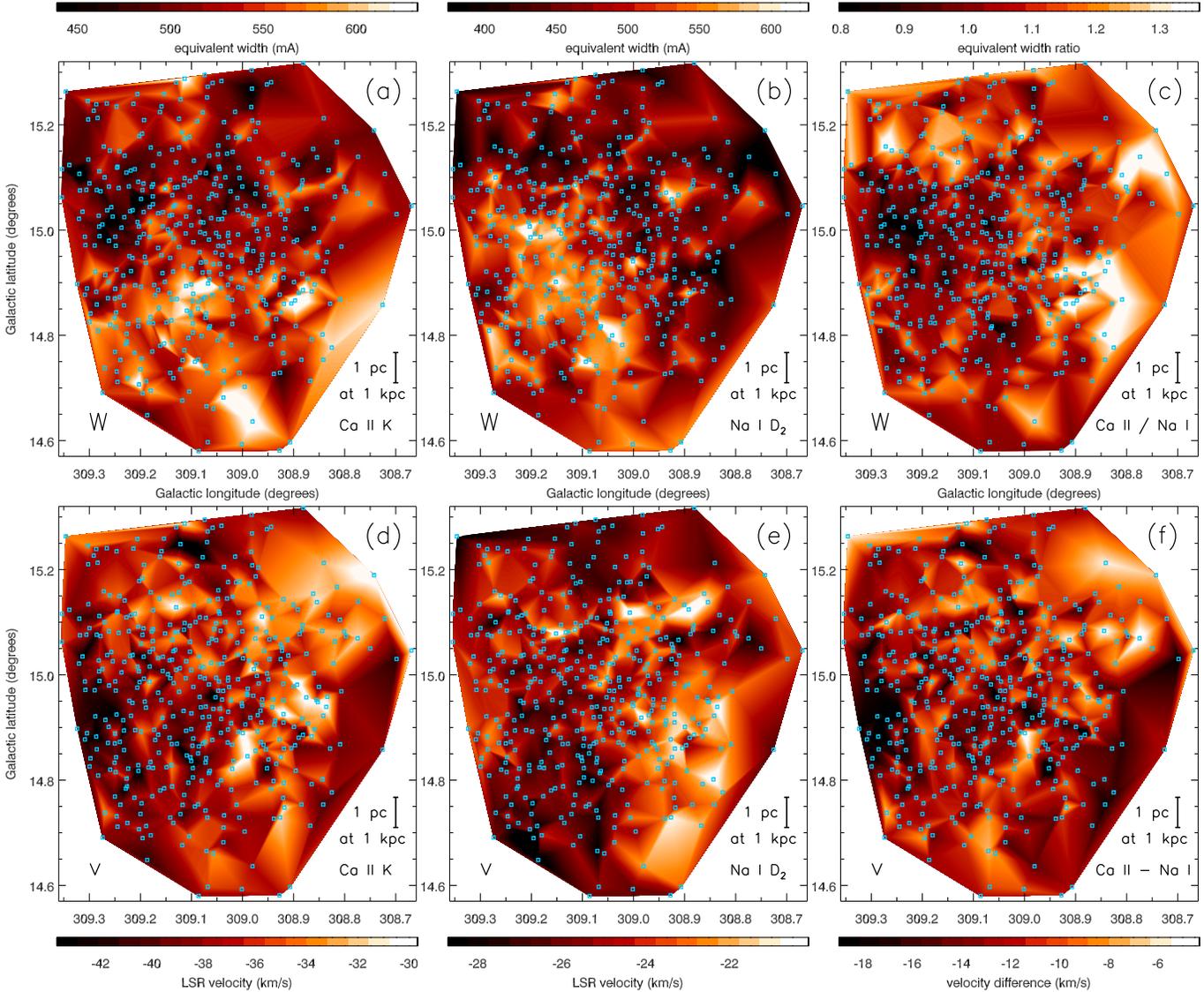,width=180mm}}
\caption[]{Maps of the equivalent width ({\it top row}) and mean velocity
({\it bottom row}) of the interstellar Ca\,{\sc ii} K ({\it left}) and
Na\,{\sc i} D$_2$ ({\it middle}) lines, and their ratio or difference,
respectively ({\it right}). The levels vary between $\pm2$ standard deviations
around the median (high values of absorption equivalent width and hence column
densities are light-coloured). The positions of the stellar probes are plotted
too.}
\end{figure*}

The equivalent widths, line widths and central velocities of the interstellar
Ca\,{\sc ii} K and Na\,{\sc i} D$_2$ lines were measured by fitting a Gaussian
(Table 1). This worked satisfactorily, the average result of which can be
judged in Fig.\ 4. Note that the line width is given in terms of the $\sigma$
value of the Gaussian, and hence is related to the Full-Width at Half Maximum
as $FWHM=2(2 \ln 2)^{1/2}\sigma=2.355\,\sigma$. This includes the contribution
of the instrumental broadening, $\sigma_{\rm i}\sim15.5$ and $\sim17.6$
km~s$^{-1}$ for the Ca\,{\sc ii} K and Na\,{\sc i} D$_2$ lines, respectively.
The formal errors on the velocities are a tenth of the spectral resolution
element, or $\sim4$ km~s$^{-1}$(\footnote{A check on the photospheric
components reveals coincidence between Ca\,{\sc ii} and Na\,{\sc i} lines in
$\omega$\,Cen stars within a few km~s$^{-1}$.}). The errors on the equivalent
widths were derived from the standard deviation of the residuals of the
Gaussian fitting, summed in quadrature and weighted by the Gaussian fit.
Typical errors are $\sim$20--30 m\AA\ for the Ca\,{\sc ii} K line and
$\sim$15--20 m\AA\ for the Na\,{\sc i} D$_2$ line.

The equivalent widths in $\sim450$ sightlines were then used to construct maps
of the interstellar absorption as well as the line ratio (Fig.\ 5). The
equivalent widths of both the Ca\,{\sc ii} K and Na\,{\sc i} D$_2$ lines
varied very significantly compared to the precision of the measurements, but
by less than a factor two across the maps (Fig.\ 5a and b): $\langle W_{\rm
Ca\,II\,K}\rangle=536$ m\AA\ (range 383--709 m\AA); $\langle W_{\rm
Na\,I\,D2}\rangle=500$ m\AA\ (range 350--682 m\AA). The standard deviations in
the distributions of $W_{\rm Ca\,II\,K}$ and $W_{\rm Na\,I\,D2}$ values are
48 and 63 m\AA, or 9\% and 13\%, respectively, compared to the mean errors in
the equivalent width values of 25 and 15 m\AA, or 5\% and 3\%, respectively.
The real fluctuations in the Ca\,{\sc ii} K and Na\,{\sc i} D$_2$ maps thus
amount to $\sim7$\% and 12\%, respectively.

The equivalent width values can be converted to column densities in the lower
energy level of the transition, following van Dishoeck \& Black (1989):
\begin{equation}
N = 1.13\times10^{17}\times\frac{W}{f\lambda^2},
\end{equation}
where the wavelength is in \AA, the equivalent width is in m\AA, and the
oscillator strengths are $f=0.65$ and 0.67 for the Ca\,{\sc ii} K and Na\,{\sc
i} D$_2$ lines, respectively. We thus obtain $\langle N_{\rm
Ca\,II\,K}\rangle\approx6.0\times10^{12}$ cm$^{-2}$; $\langle N_{\rm
Na\,I\,D2}\rangle\approx2.4\times10^{12}$ cm$^{-2}$. The conversion to column
density assumes that the lines are weak enough to be on the linear part of the
curve-of-growth. Given the moderate values for the equivalent width and
reddening, E(B--V), this is not a bad assumption (see Crawford 1992a for
examples of resolved line profiles at different E(B--V) values). It also
assumes a single kinematic component, whereas in our case the absorption is a
blend of at least two kinematically distinct components (see below).

Structure is visible on all scales, between neighbouring sightlines (but not
always) as well as across the $0.7^\circ$ field. The atomic line ratio varied
slightly more (Fig.\ 5c), $\langle W_{\rm Ca\,II\,K}/W_{\rm
Na\,I\,D2}\rangle=1.08$ (range 0.72--1.54), or $\langle N_{\rm
Ca\,II\,K}/N_{\rm Na\,I\,D2}\rangle\approx2.5$ (range 1.7--3.6). Indeed,
although there are regions where the two lines vary in concert (e.g., the
lower-left area and upper border in Fig.\ 5), there are regions where either
the Ca\,{\sc ii} K (e.g., the top-left corner and right border in Fig.\ 5) or
the Na\,{\sc i} D$_2$ (Fig.\ 5: around $l=309.25^\circ$, $b=15.0^\circ$)
absorption is relatively strong.

The Na\,{\sc i} D$_2$ line is broader in wavelength than the Ca\,{\sc ii} K
line, but not in velocity: $\langle\sigma_{\rm Na\,I\,D2}\rangle=23$
km~s$^{-1}$ (range 19--30 km~s$^{-1}$), whereas $\langle\sigma_{\rm
Ca\,II\,K}\rangle=30$ km~s$^{-1}$ (range 24--37 km~s$^{-1}$). Interestingly,
the ratio of equivalent widths correlates significantly with that of the line
widths (Fig.\ 6a, with Pearson's values $r=0.37$, $p<0.0001$), and a
corresponding shift in the central velocity between the two lines is seen
(Fig.\ 6b, $r=-0.31$, $p<0.0001$). This suggests that at least some sightlines
pass through multiple absorption components (clouds), with Na$^0$ clouds
appearing at somewhat more positive velocities, and Ca$^+$ clouds at more
negative velocities. We will come back to this observation in the discussion
section.

%
%
\begin{figure}
\centerline{\psfig{figure=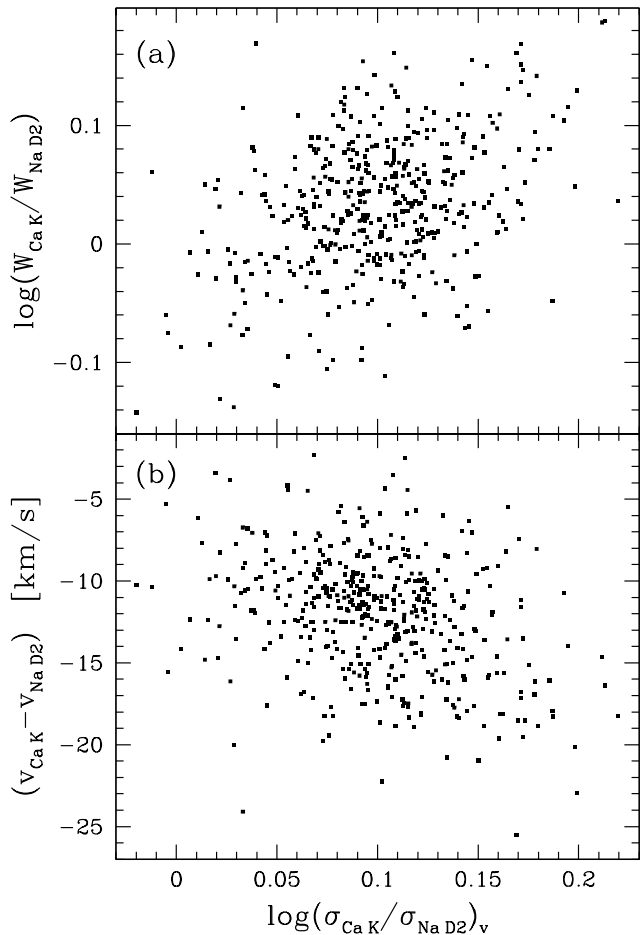,width=84mm}}
\caption[]{Comparison between interstellar Ca\,{\sc ii} K and Na\,{\sc i}
D$_2$ lines: equivalent widths ({\it a}) and velocities ({\it b}) {\it versus}
the $\sigma$ of the Gaussian fits expressed in velocity.}
\end{figure}

\subsection{Diffuse Interstellar Bands}

%
%
\begin{figure}
\centerline{\psfig{figure=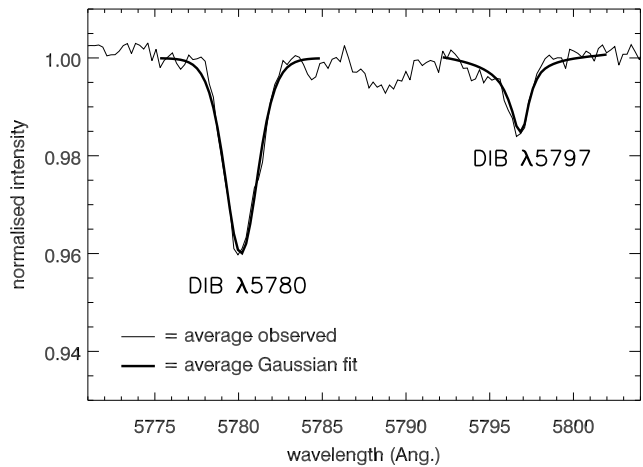,width=84mm}}
\caption[]{The average spectrum around the $\lambda$5780 and $\lambda$5797
DIBs. The average Gaussian fit is overlain.}
\end{figure}

The first DIB features to have been found (Heger 1922) and shown to be of
interstellar origin (Merrill 1934) are at 5780 and 5797 \AA. The modest
reddening towards our stellar probes not surprisingly results in rather weak
DIB features, but in the average spectrum they are clearly detected at 4 and
2\% depth with respect to the continuum (Fig.\ 7).

Gaussian fitting was on the whole still adequate, with typical errors of
$\sim$20--40 m\AA\ for the $\lambda$5780 DIB and $\sim$15--25 m\AA\ for the
weaker but sharper $\lambda$5797 DIB. Only fits of reasonable fidelity were
kept by imposing a threshold on the equivalent width, $>20$ m\AA, and
constraining the bandwidth ($\sigma$) to within 0.15--1.8 \AA\ for the
$\lambda$5780 band and 0.15--1 \AA\ for the $\lambda$5797 band. Thus we kept
the majority (415) of $\lambda$5780 fits but only a third (164) of the
$\lambda$5797 fits. For these, the equivalent widths were $\langle W_{\rm
DIB\,5780}\rangle=100$ m\AA, up to a maximum of 251 m\AA, with a bandwidth
$\langle\sigma_{\rm DIB\,5780}\rangle=0.95$ \AA, and $\langle W_{\rm
DIB\,5797}\rangle=33$ m\AA, up to a maximum of 128 m\AA, with a bandwidth
$\langle\sigma_{\rm DIB\,5797}\rangle=0.49$ \AA. The shape of the
$\lambda$5797 band deviates from that of a Gaussian profile, not least due to
a broader, shallower band at 5795 \AA. Considering the weakness of the
$\lambda$5797 band in our spectra the Gaussian fitting is as accurate as any
method.

%
%
\begin{figure}
\centerline{\psfig{figure=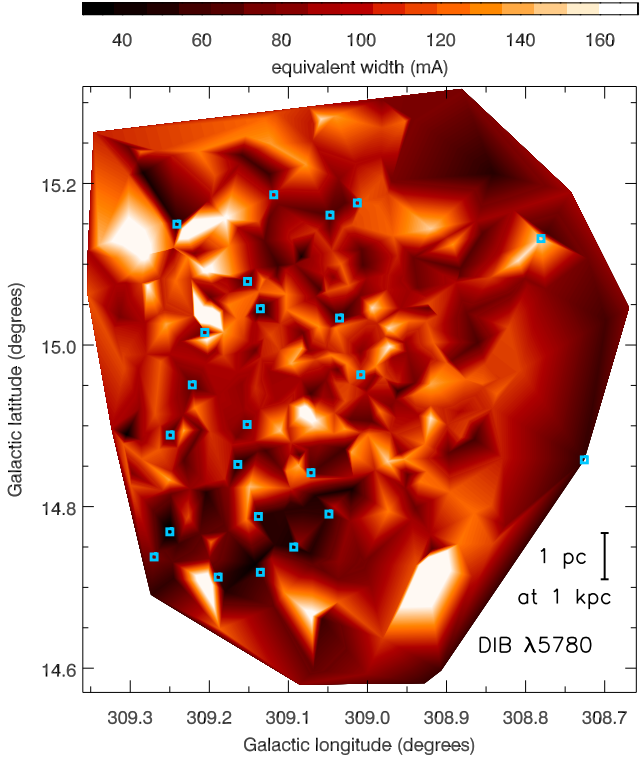,width=84mm}}
\caption[]{Map of the equivalent width of the $\lambda$5780 DIB. The intensity
levels vary between $\pm2$ standard deviations around the median. Positions of
the stellar probes are plotted only for sightlines with $\log(W_{\rm
DIB\,5797}/W_{\rm DIB\,5780})>-0.2$.}
\end{figure}

There are substantial variations of the DIB strength across the $0.7^\circ$
field and these are mapped in the stronger $\lambda$5780 DIB in Fig.\ 8. This
DIB is generally stronger where the Ca\,{\sc ii} K line is weaker (compare
Fig.\ 5). However, some strong $\lambda$5780 absorption is seen in the upper
left where the Na\,{\sc i} D$_2$ is weak but the Ca\,{\sc ii} K line is not
weak. Despite fewer sightlines the DIB map appears to have more small-scale
structure, possibly because DIB absorption is generally associated with the
neutral cloud medium which has a smaller filling factor than the warmer
inter-cloud medium. However, much of the scatter is due to the uncertainties
in the equivalent width measurements. The mean error in $W_{\rm DIB\,5780}$ is
32 m\AA, whilst the standard deviation of the distribution of $W_{\rm
DIB\,5780}$ values (i.e.\ the fluctuations in the map) is 35 m\AA. Given the
mean value for $W_{\rm DIB\,5780}$ of 100 m\AA, this would suggest that the
real fluctuations amount to ``only'' $\sim15$ m\AA (15\%), still comparable to
--- or exceeding --- the fluctuations in Ca\,{\sc ii} K and Na\,{\sc i} D$_2$.

A significant positive correlation exists for the ratio of $\lambda$5780 DIB
and Na\,{\sc i} D$_2$ equivalent widths and the ratio of Ca\,{\sc ii} K and
Na\,{\sc i} D$_2$ equivalent widths (Fig.\ 9b, $r=+0.26$, $p<0.0001$; very
small adjustments were made to the r-values to account for correlated errors
in $W_{\rm Ca\,II\,K}$ and $W_{\rm Na\,I\,D2}$). This suggests that this DIB
is strong where Ca$^+$ is more abundant. There is no significant correlation
between the ratio of $\lambda$5780 DIB and Ca\,{\sc ii} K equivalent widths
and the Ca\,{\sc ii} K/Na\,{\sc i} D$_2$ ratio (Fig.\ 9a, $r=-0.06$,
$p=0.24$). There is no clear trend of the DIB ratio with Ca\,{\sc ii}
K/Na\,{\sc i} D$_2$ ratio either (Fig.\ 9c), but if the strongest $\zeta$-type
absorption is selected (see below), $\log(W_{\rm DIB\,5797}/W_{\rm
DIB\,5780})>-0.2$, then these sightlines tend to be where the $\lambda$5780
DIB is weak, but both Ca\,{\sc ii} K and Na\,{\sc i} D$_2$ absorption are
fairly strong (Figs.\ 5 \& 8).

%
\begin{figure}
\centerline{\psfig{figure=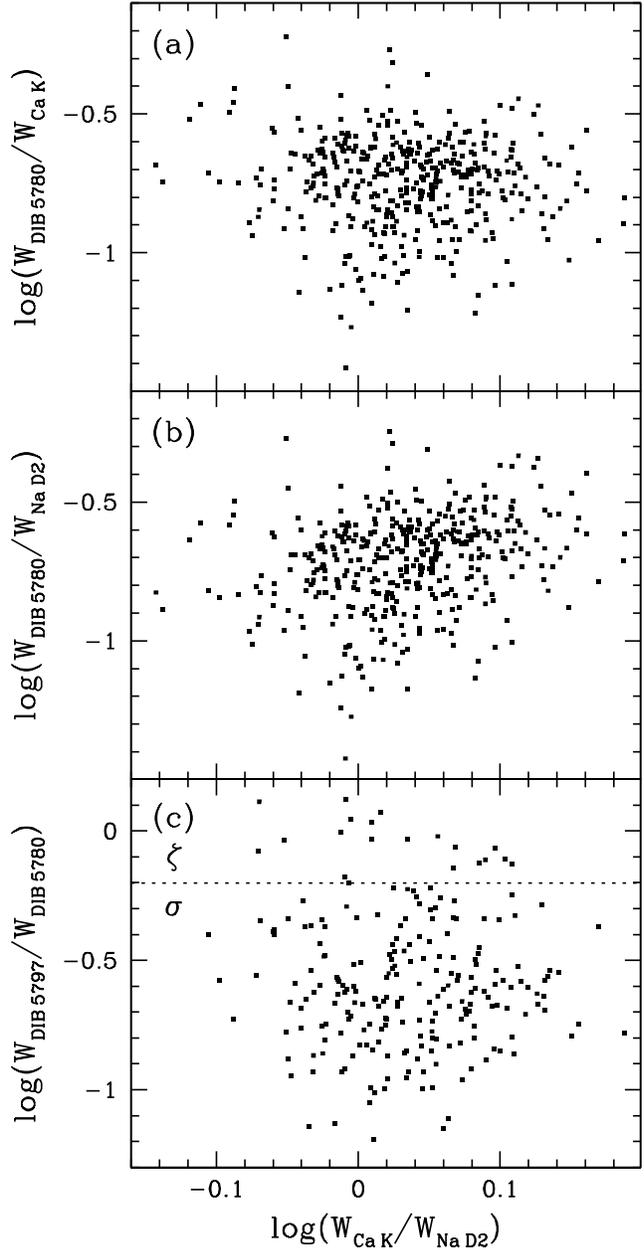,width=84mm}}
\caption[]{Comparison of the equivalent widths of the $\lambda$5780 and
$\lambda$5797 DIBs, and the interstellar Ca\,{\sc ii} K and Na\,{\sc i} D$_2$
lines.}
\end{figure}

The $W_{\rm DIB\,5797}/W_{\rm DIB\,5780}$ ratio is an indicator of the
ionization conditions as inferred from the atomic to total (H + H$_2$)
hydrogen ratio (Cami et al.\ 1997). The $\lambda$5780 DIB is relatively weak
in $\zeta$-Per type clouds and strong in $\sigma$-Sco type clouds
(Kre{\l}owski \& Sneden 1995), so the Disc--Halo interface region appears to
be of $\sigma$ type. The charge state of the $\lambda$5780 and $\lambda$5797
band carriers is not known, but if both are, for example, singly charged
cations, then the low-UV ($\zeta$) c.f.\ high-UV ($\sigma$) classification is
consistent with the parent neutrals having low ($\lambda$5797) and relatively
high ($\lambda$5780) first photo-ionization energies such that the
$\lambda$5780 absorption appears when the UV field is strong. The
$\sigma$-type character observed towards $\omega$\,Cen is consistent with
other observations of the $\lambda$5797 and $\lambda$5780 diffuse bands along
sightlines out of the Galactic plane. Examples of observations of Milky Way
diffuse band absorption are those seen towards the Magellanic Clouds
(Ehrenfreund et al.\ 2002; Welty et al.\ 2006), and also SN\,1987A (LMC)
(Vladilo et al.\ 1987) and SN\,1986G (NGC\,5128) (D'Odorico et al.\ 1989) for
which the reported $W_{\rm DIB\,5797}/W_{\rm DIB\,5780}$ ratios are 1/4 and
1/2, respectively.

The $\lambda$6196 and $\lambda$6203 DIBs are also seen in some of the spectra.
However, this is close to the edge of our spectral range, and is in fact not
covered in all spectra. In the spectra that do cover these bands, they are
generally too faint to allow a meaningful analysis.

\subsection{Evidence for gas associated with $\omega$\,Centauri?}

The spectral resolution element exceeds the internal velocity dispersion of
the $\omega$\,Cen cluster (e.g., van Loon et al.\ 2007), and the generally
weak but present photospheric absorption in both the Ca\,{\sc ii} H and K
lines and the Na\,{\sc i} D$_1$ and D$_2$ lines further limits searches for
absorption components arising in the intra-cluster medium (ICM) within
$\omega$\,Cen. One could only hope to find very strong absorption in
particularly hot stars, or, more sensitively, significantly displaced
absorption because of the stellar peculiar velocity with respect to the
cluster's systemic velocity.

The average spectrum does not reveal any absorption in addition to the local
interstellar components around the Sun's velocity and photospheric components
of the stellar probes in $\omega$\,Cen (Fig.\ 4). If large variations were to
occur in the ICM, then there might be a possibility to discover ICM components
in individual spectra. However, the signal-to-noise ratio of the individual
spectra is limited, and the resulting column density sensitivity is
$N\,\gsim\,10^{11}$ cm$^{-2}$ for the Ca\,{\sc ii} H and K and Na\,{\sc i}
D$_1$ and D$_2$ lines. No evidence was found for ICM at a level exceeding
this. Assuming a uniform column density across a circular area of
$R=0.3^\circ$ radius, this corresponds to a limit of $M_{\rm ICM}\,\lsim\,4$
M$_\odot$ unless Na$^0$ and Ca$^+$ were to be grossly under-representative
species. This estimate is consistent with the $M_{\rm ICM}\,\lsim\,3$
M$_\odot$ from {\it Spitzer} maps of $\omega$\,Cen (Boyer et al.\ 2008;
McDonald et al.\ 2009), and reinforces evidence for efficient clearing of the
cluster by interaction with the hot Halo (van Loon et al.\ 2009).

\section{Discussion}

\subsection{Density, ionization, or dust depletion?}

Are we probing density variations, or fluctuations in ionization and
excitation equilibrium, e.g., due to variations in the interstellar radiation
field? The Solar abundances of sodium and calcium are nearly identical, and
hence the relative column densities of Ca$^+$ and Na$^0$ readily indicate the
dust depletion and excitation/ionization conditions. Purely
thermally-broadened lines of sodium and calcium are unresolved in our spectra.
With a mean atomic weight of 23 and 40, respectively, the line width would be
$\sigma<19$ km s$^{-1}$ even for a $10^6$ K plasma. Resolution of these lines
therefore signifies the presence of multiple clouds along the sightline,
rather than providing a means to estimate the gas kinetic temperature.
Furthermore, the ionization state is likely to be dominated by the local
strength of the interstellar radiation field, and thus electron density,
neither of which are particularly well-constrained.

%
\begin{figure*}
\centerline{\hbox{
\psfig{figure=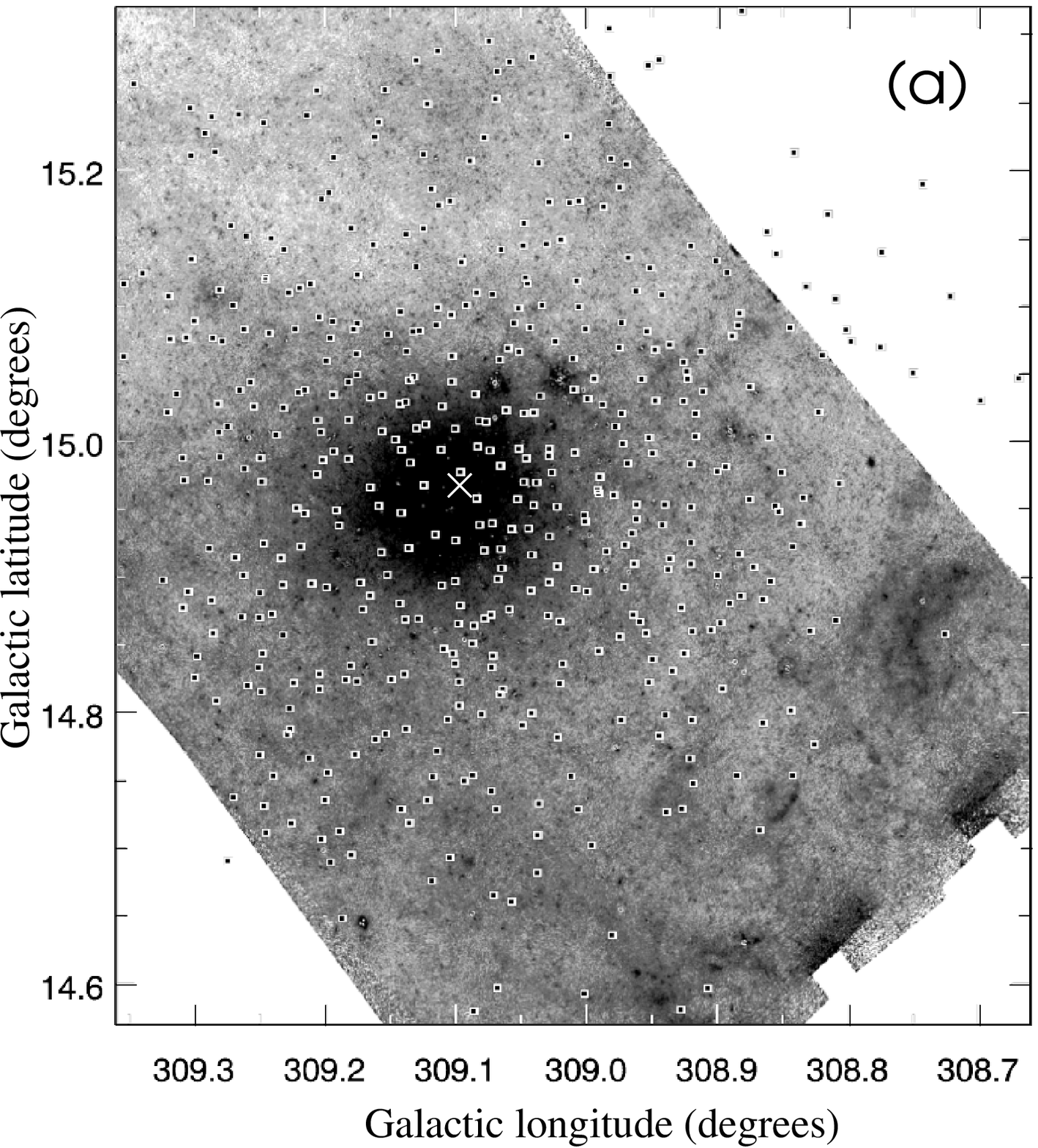,width=86mm}
\psfig{figure=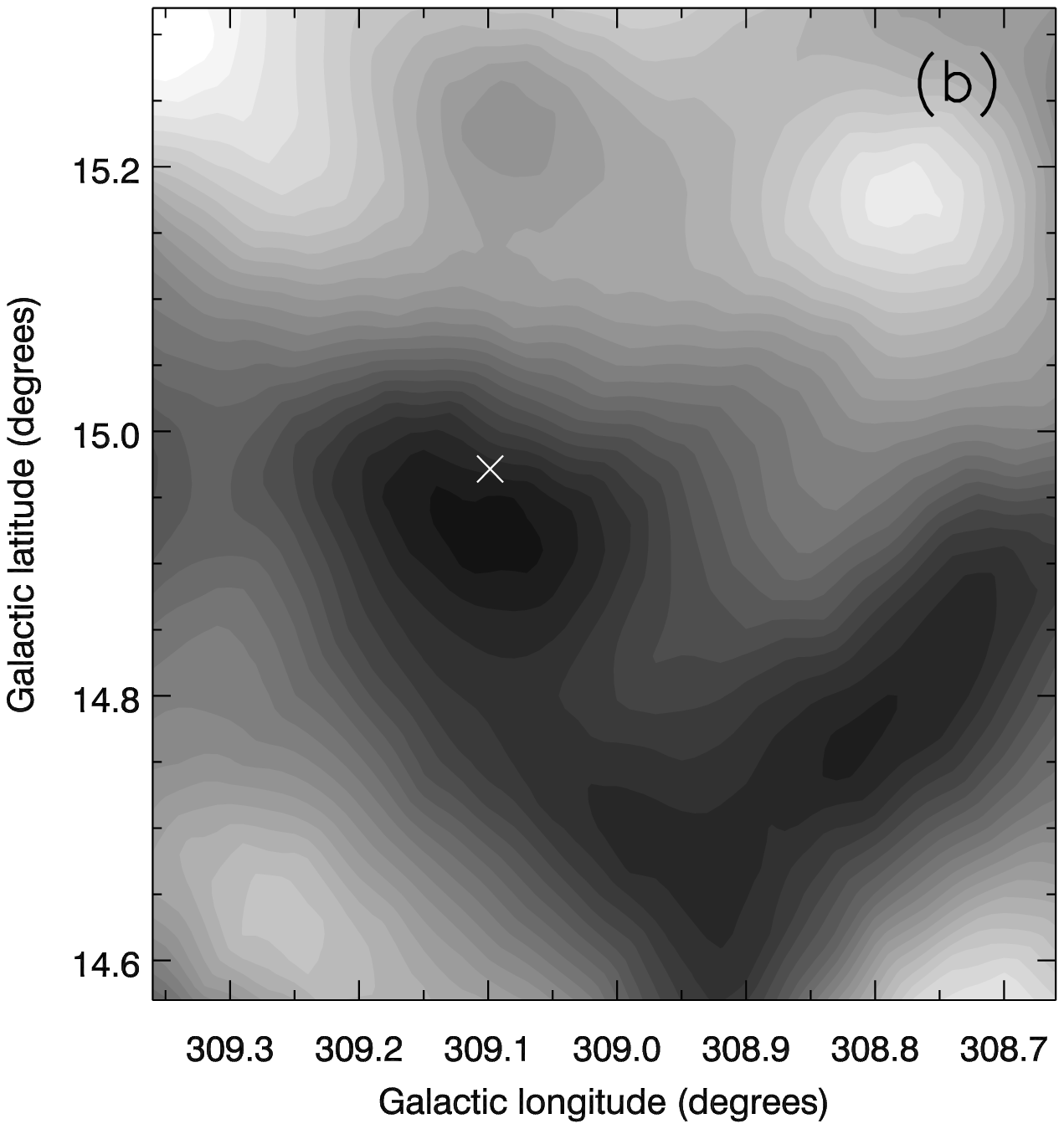,width=88mm}
}}
\caption[]{({\it a}:) {\it Spitzer Space Telescope} 24-$\mu$m image (c.f.\
Boyer et al.\ 2008) which, besides starlight from $\omega$\,Cen, shows warm
dust. The sightlines of the absorption-line experiment are overplotted. ({\it
b}:) Map of the reddening, derived from DIRBE/IRAS data (Schlegel et al.\
1998). Values range from $E(B-V)=0.12$ (white) to 0.144 mag (dark). In both
panels the centre of $\omega$\,Cen is indicated by a cross.}
\end{figure*}

One may assume that the occupancies of the 3s and 4s states for the Na\,{\sc
i} D and Ca\,{\sc ii} doublets are mostly related to the ionization state
rather than a different degree of excitation of a particular atomic species.
The relevant ionization potentials are as follows: Ca$^0$, 6.11 eV; Ca$^+$,
11.87 eV; Ca$^{++}$, 50.91 eV; Na$^0$, 5.14 eV; Na$^+$, 47.29 eV. This means
that in the warm ionized medium (significant energy density above 13.6 eV),
neither Na$^0$ nor Ca$^+$ are the dominant species. Still, these lower
ionization stages remain well populated and, given the high ionization
potentials of the next higher stage (in both cases), should be fairly stable
in their occupancy. In the cold neutral medium (with H$_2$), Na$^0$ is
dominant but there is no Ca$^+$. Therefore, no correspondence between Ca\,{\sc
ii} and Na\,{\sc i} can be found in the cold neutral medium. In the warm
neutral/weakly-ionized medium (atomic H), Ca$^+$ is abundant, but perhaps more
susceptible to variations in radiation field than Na$^0$ given that both
ionizations to Ca$^{++}$ and recombinations to Ca$^0$ are frequent (c.f.\
Crawford 1992b).

The strong Na\,{\sc i} 8183+8195 \AA\ doublet corresponds to the transition
upwards from the next state, 3p, but our spectra do not include this line. The
Ca\,{\sc i} 4227 \AA\ line corresponds to the equivalent transition to the
Ca\,{\sc ii} doublet but in the neutral phase, but this line is not covered by
our spectra either. The 4227 \AA\ line is extremely weak or absent in earlier
2dF spectra (Figs.\ 6 \& 11 in van Loon 2002). Careful re-examination of the
2dF spectra of the few hundred hottest stars in van Loon et al.\ (2007)
resulted in a detection rate of a few per cent at most, with equivalent widths
not exceeding a few per cent of that of the Ca\,{\sc ii} K line. Thus we have
no direct means of constraining the Boltzmann-Saha distributions over the
electronic energy levels.

\subsubsection{Sodium and calcium as tracers of dust}

Calcium is a highly refractive element and thus easily depleted from the gas
phase onto dust grains, even in diffuse clouds. Sodium is much less affected,
which allows a test to be made by comparing the Na\,{\sc i} and Ca\,{\sc ii}
equivalent widths with the reddening, E(B--V).

First we compare our absorption-line maps with the partly-overlapping {\it
Spitzer Space Telescope} 24-$\mu$m map (Fig.\ 10; c.f.\ Boyer et al.\ 2008).
The latter shows starlight from the cluster in the central portion of the
image, but mostly indicates where relatively warm dust is located. Focussing
on the band of low 24-$\mu$m surface brightness running from about
$l=308.9^\circ$, $b=15.1^\circ$ to about $l=309.3^\circ$, $b=15.2^\circ$, this
is where the Na\,{\sc i} absorption is weak but the ${\rm W}_{\rm
Ca\,II\,K}/{\rm W}_{\rm Na\,I\,D2}$ ratio is fairly high. This suggests that
we see warm material with little sign of calcium depletion, as expected in
regions with little dust. Just below this band, around
$l=309.2$--$309.3^\circ$, $b=15.0^\circ$, lies of patch of dust emission; this
is where Na\,{\sc i} is strong and the ${\rm W}_{\rm Ca\,II\,K}/{\rm W}_{\rm
Na\,I\,D2}$ ratio reaches the lowest values in our maps, suggesting high
column densities of the neutral medium with strong calcium depletion.

To better quantify the effects of depletion and test how well Na\,{\sc i}
traces the neutral medium, we correlate the equivalent widths with reddening
values taken from the DIRBE/IRAS maps of cold dust (Schlegel, Finkbeiner \&
Davis 1998). These vary by $\sim20$\% across the maps, between
$E(B-V)\simeq0.120$--0.144 mag; the reddening values are precise to 16\%
(Schlegel et al.\ 1998), although the statistical spread must be considerably
smaller within the confines of this small portion on the sky (c.f.\ Fig.\
10b). In spite of the vastly different angular resolution, the reddening map
(Fig.\ 10b) reveals clear parallels with the {\it Spitzer} 24-$\mu$m map
(Fig.\ 10a). For instance, there is a dense cloud stretching across the
lower-right corner, and a general lack of dust in the top part of the region.
Curiously, the reddening map suggests a dust cloud very near (but just below)
the centre of $\omega$\,Cen. This might have been affected by far-IR starlight
from the cluster, but the Na\,{\sc i} absorption (in particular) is also
rather strong in that patch of sky (Fig.\ 5).

%
\begin{figure}
\centerline{\psfig{figure=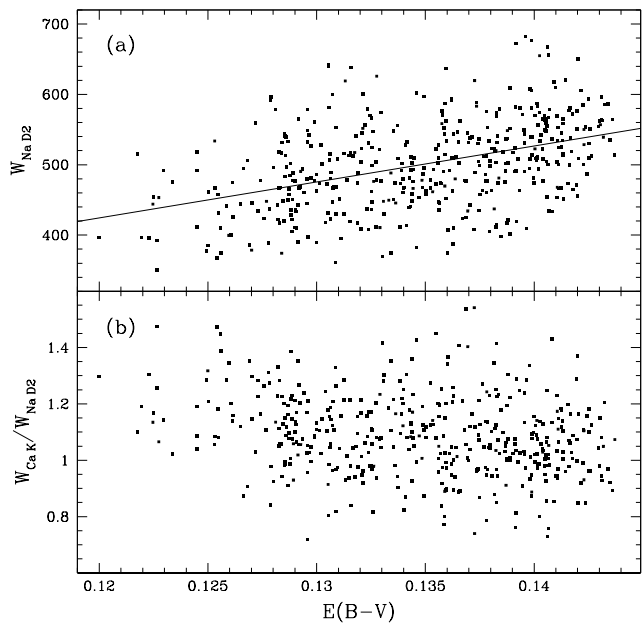,width=84mm}}
\caption[]{Comparison of the equivalent widths of the interstellar Ca\,{\sc
ii} K and Na\,{\sc i} D$_2$ lines as a function of E(B--V) derived from
DIRBE/IRAS data, with a linear fit to $W_{\rm Na\,I\,D2}$ {\it versus}
E(B--V).}
\end{figure}

The clearest correlation with E(B--V) is for Na\,{\sc i} (Fig.\ 11a,
$r=+0.43$, $p<0.0001$), which increases in strength by at least $\sim40$\%.
This confirms that Na\,{\sc i} indeed traces the column density of the neutral
dusty medium  (Hobbs 1974). The fact that it seems to increase more strongly
than the reddening does, can most simply (and hence plausibly) be explained by
a need to re-adjust the reddening values from the Schlegel et al.\ (1998) maps
towards lower values. Indeed, the mean $\langle E(B-V)\rangle\sim0.13$ mag is
$\sim0.05$ mag higher than deduced from accurate analysis of the
colour--magnitude diagram of stars within $\omega$\,Cen (McDonald et al.\
2009). If the correction is a simple offset, leaving the spread intact, then
the reddening would vary by more than 20\%, possibly by about 40\% just as the
Na\,{\sc i} line strength does. Although McDonald et al.\ (2009) constrain
differential reddening gradients across the cluster to $\Delta
E(B-V)\,\lsim\,0.02$ mag near the cluster centre, Calamida et al.\ (2005)
inferred reddening variations up to a factor almost two from the colours of
hot HB stars in $\omega$\,Cen, indicating a clumpy medium; given that the
Schlegel et al.\ maps have a rather coarse resolution some of the spread in
the Na\,{\sc i} line strength may be due to such clumps.

The ${\rm W}_{\rm Ca\,II\,K}/{\rm W}_{\rm Na\,I\,D2}$ ratio shows a weak
anti-correlation with E(B--V), at the $\sim20$\% level (Fig.\ 11b, $r=-0.24$,
$p<0.0001$), consistent with the effects of dust depletion. However, given
that the Na\,{\sc i} line strengthens even more, it means that the Ca\,{\sc
ii} still appears generally stronger along dustier sightlines and therefore,
to some extent, also traces the column density of the neutral medium (c.f.\
Hobbs 1974).

%
\begin{figure}
\centerline{\psfig{figure=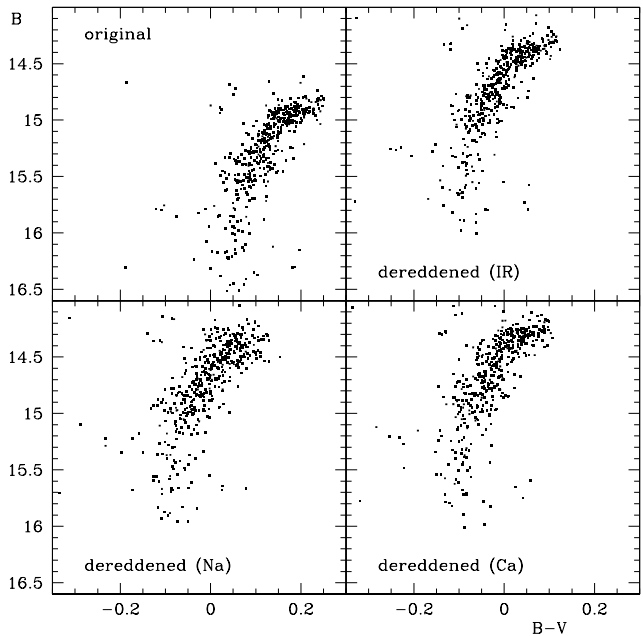,width=84mm}}
\caption[]{B, B--V diagram of the spectroscopic targets in $\omega$\,Centauri
(top left), and dereddened using the Schlegel et al.\ (1998) DIRBE/IRAS map
(top right), or a parameterisation of E(B--V) with Na\,{\sc i} D$_2$ (bottom
left) or with Ca\,{\sc ii} K (bottom right).}
\end{figure}

To test whether the Na\,{\sc i} or Ca\,{\sc ii} lines could be used to trace
the small-scale distribution of interstellar dust, we deredden the B, B--V
diagram using the DIRBE/IRAS maps, and again using instead a scaling of the
reddening with the Na\,{\sc i} and Ca\,{\sc ii} equivalent width (Fig.\ 12).
We parameterised $E(B-V)={\rm W}_{\rm Na\,I\,D2}[{\rm \AA}]/3.8$ and
$E(B-V)={\rm W}_{\rm Ca\,II\,K}[{\rm \AA}]/3.6$, and assumed $A_B=4.07
E(B-V)$.

Remarkably, the coarser DIRBE/IRAS map is much better in dereddening the
colour--magnitude diagram than the absorption lines, despite the latter being
observed along exactly the same sightlines as the stars that are being
corrected. The clearest evidence for this statement is that the scatter in the
HB sequence increases significantly when using the absorption lines, whereas
it remains as tight as the original, observed sequence when using the
DIRBE/IRAS maps. The errors in the equivalent width are typically
$\lsim\,5$\%, translating into errors in E(B--V) and A$_{\rm B}$ of
$\lsim\,0.01$ and $\lsim\,0.05$ mag, respectively, i.e.\ smaller than the
increased scatter. The increased scatter, especially when using the Na\,{\sc
i} absorption, would tend to imply that the variations arise from fluctuations
in ionization equilibrium rather than total atomic column density, assuming
that the atomic column densities are well correlated with dust. In any case it
confirms that the fluctuations in the Ca\,{\sc ii} K and Na\,{\sc i} D$_2$
lines are a physical reality.

\subsubsection{The Diffuse Interstellar Bands}

The $\lambda$5780 and $\lambda$5797 DIBs are encountered in diffuse clouds,
with the $\lambda$5780 DIB appearing stronger where the UV radiation field is
stronger ($\sigma$-type clouds) --- which presumably leads to the destruction
of the carrier of the $\lambda$5797 DIB which is stronger in denser neutral
media ($\zeta$-type clouds) (Kre{\l}owski \& Westerlund 1988; Kre{\l}owski et
al. 1999). If both are singly-charged macromolecules, then their different UV
threshold for appearance might be related to differences in the
photo-ionization energy of their parent neutral molecules, that of the 5780
\AA\ DIB being higher (c.f.\ Cami et al.\ 1997). Alternatively, it might
simply reflect a difference in charge, with the $\lambda$5780 DIB due to a
weakly ionized or protonated macromolecule and the $\lambda$5797 DIB due to a
neutral macromolecule, much like the variations suggested in the mid-IR
spectra of Polycyclic Aromatic Hydrocarbons (Allamandola, Tielens \& Barker
1989).

The sightlines we probe here would be more translucent than in the Galactic
plane, and hence the $\sigma$ behaviour would be expected to dominate. Indeed,
few sightlines in our data have a strong $\lambda$5797 DIB, consistent with a
low filling factor of the cold neutral medium. The highest ${\rm W}_{\rm
DIB\,5797}/{\rm W}_{\rm DIB 5780}$ ratios are seen where the $\lambda$5780
band is weak not because the $\lambda$5797 band is strong (Fig.\ 8). But more
typically, it is the $\lambda$5797 band which varies: in Fig.\ 13 two spectra
are compared, showing weak and strong $\lambda$5797 bands for very similar
$\lambda5780$ bands and Ca\,{\sc ii} and Na\,{\sc i} lines. The former is an
average of the spectra of LEID\,33137, 37193, 40185, 41087, 42153, 64042 and
75026, whilst the latter is an average of the spectra of LEID\,20006, 21022,
26096, 55048, 60020 and 68026.

%
\begin{figure}
\centerline{\vbox{
\psfig{figure=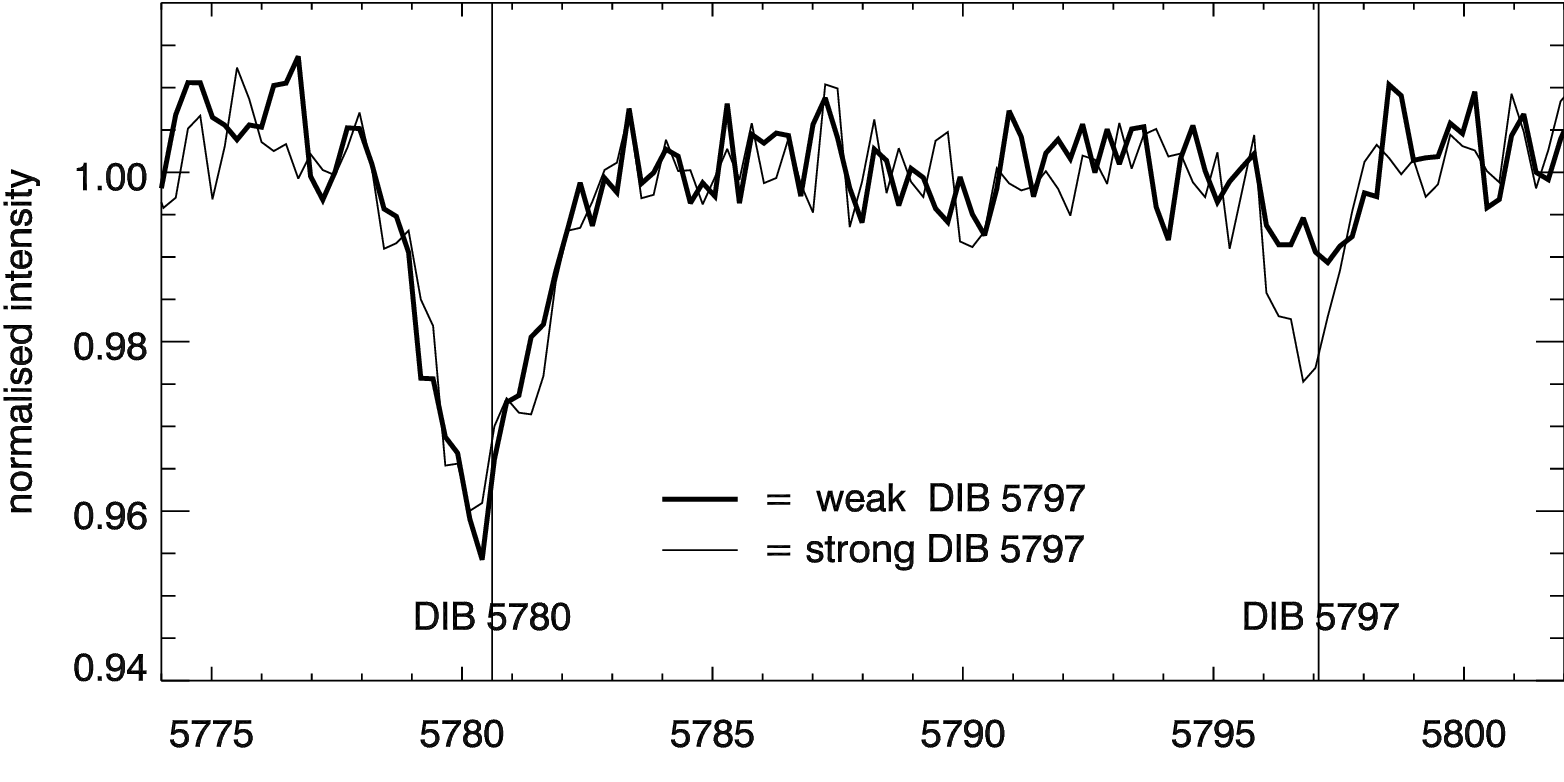,width=84mm}
\psfig{figure=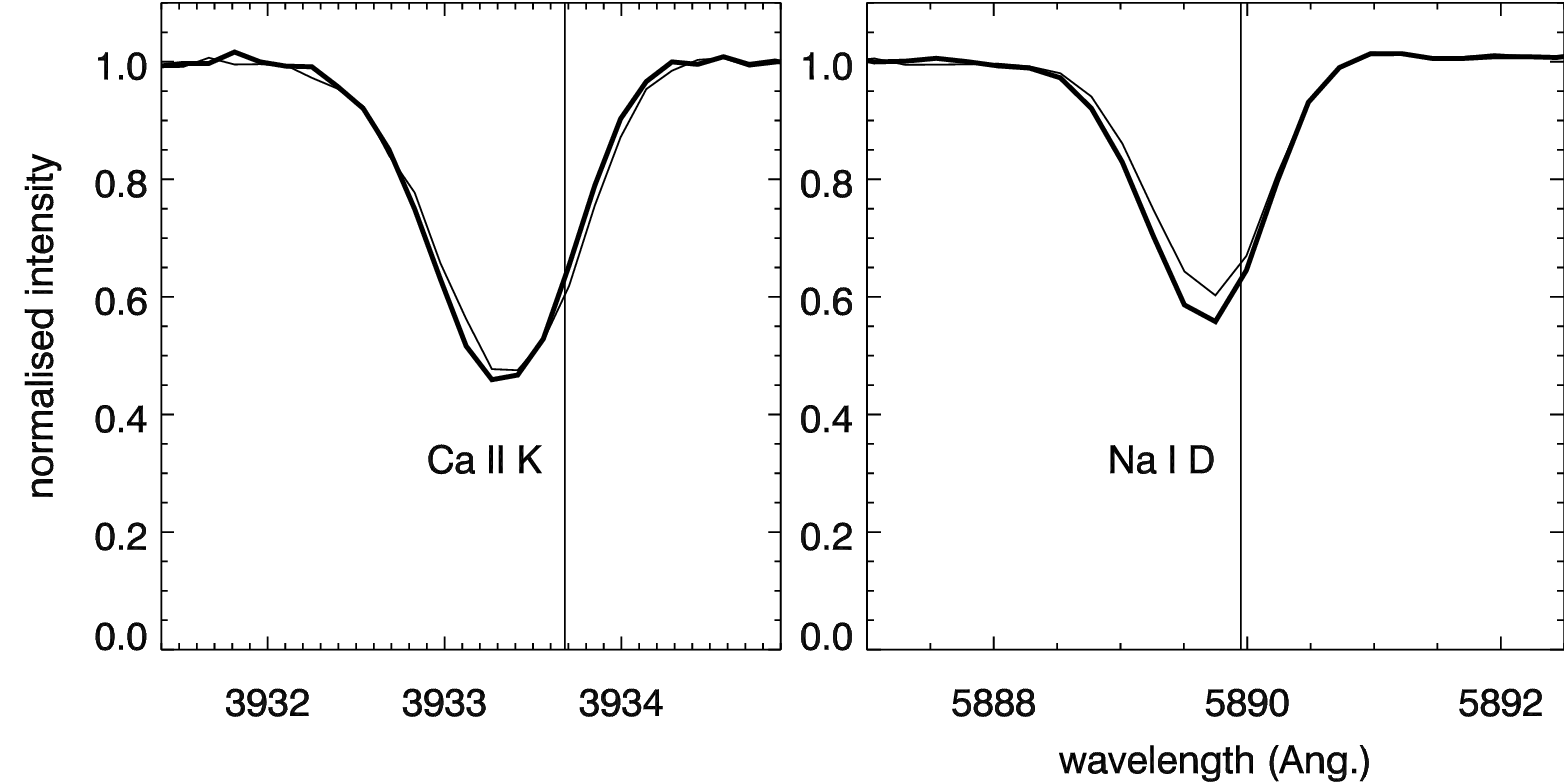,width=84mm}
}}
\caption[]{Comparison between two average spectra, differing mainly in the
strength of the $\lambda$5797 DIB.}
\end{figure}

The $\lambda$5780 DIB shows no correlation with E(B--V) over the small range
in E(B--V) probed here. Comparing with the 24-$\mu$m map (Fig.\ 10a) there is
some correspondence between the absence of warm dust around
$l\sim309.2^\circ$, $b\sim15.15^\circ$ and stronger $\lambda$5780 DIB
absorption, and between the warm dust emission around $l\sim309.2^\circ$,
$b\sim14.7$--$14.8^\circ$ and weaker $\lambda$5780 DIB absorption. In the
latter region the $\lambda$5797 DIB is often seen, confirming the presence of
a dense medium there. There are about ten times more sightlines with $W_{\rm
DIB\,5797}/W_{\rm DIB 5780}<0.7$ than with $W_{\rm DIB\,5797}/W_{\rm DIB
5780}>0.7$, which suggests that the denser medium traced by the $\lambda$5797
DIB has a lower filling factor than the warmer neutral medium by a factor
$\sim10$.

%
\begin{figure}
\centerline{\psfig{figure=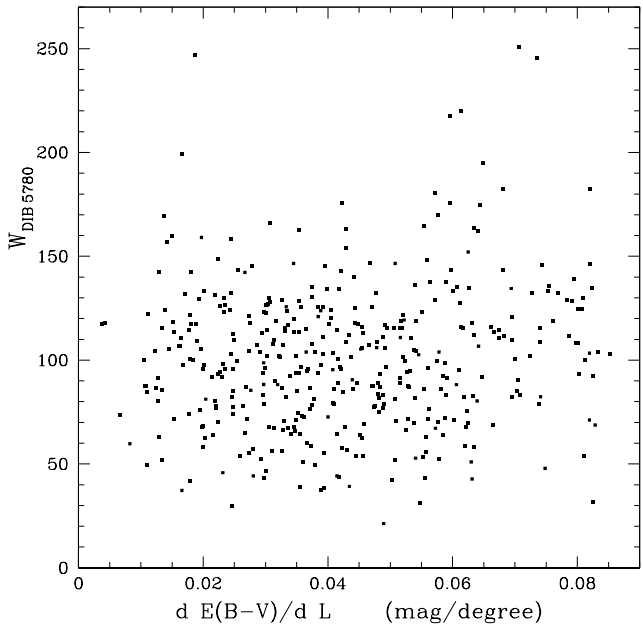,width=84mm}}
\caption[]{Correlation between the equivalent width of the $\lambda$5780 DIB
and the local maximum gradient in the DIRBE/IRAS reddening map.}
\end{figure}

The gradient in reddening can help pinpoint the edges of clouds, under the
assumption that the reddening reaches a maximum in the cloud core and a
minimum in the inter-cloud regions. The (absolute value of the) gradient would
be maximum near the boundary between cloud and inter-cloud region. There is a
very tentative indication that at the steepest gradients, ${\rm d}E(B-V)/{\rm
d}L\,\gsim\,0.06$ mag deg$^{-1}$, the $\lambda$5780 DIB is stronger than at
shallower gradients (Fig.\ 14): $\langle W_{\rm DIB 5780}\rangle_{\rm
steep}=111.4\pm4.6$ m\AA\ {\it versus} $\langle W_{\rm DIB 5780}\rangle_{\rm
shallow}=96.7\pm1.8$ m\AA, a difference of $15\pm5$ m\AA\ or 3 $\sigma$. A
positive correlation, if significant, would be consistent with the
$\lambda$5780 DIB originating in the warmer envelopes of diffuse clouds.

\subsection{The interstellar path towards $\omega$\,Centauri}

Much of the structure might be relatively near to us, seen against a smoother,
more distant background higher above the Galactic Disc. Mapping of the nearby
ISM indeed suggests that the path towards $\omega$\,Cen first traverses the
warm/hot gas of the Local Bubble until $\sim100$ pc, then passes through a
neutral medium until $\sim200$ pc before leaving the Disc (Lallement et al.\
2003). Although $\omega$\,Cen is not seen through the bright H$\alpha$
emission of the Galactic Disc, it is seen through a layer of fainter, more
diffuse H$\alpha$ emission (Gaensler et al.\ 2008) possibly associated with an
extra-planar warm ionized medium. Given that Na\,{\sc i} and Ca\,{\sc ii} both
trace the warm neutral medium one might have expected that the absorption
bears a prominent imprint of the slab of neutral cloud material.

This does not appear to be the case. The column density ratio ${\rm N}_{\rm
Ca\,II\,K}/{\rm N}_{\rm Na\,I\,D2}\sim1.7$--3.4. For warm, inter-cloud
material, Crawford (1992b) found that ${\rm N}_{\rm Ca}/{\rm N}_{\rm
Na}\approx0.036\ {\rm N}_{\rm Ca\,II\,K}/{\rm N}_{\rm Na\,I\,D2}$. In
our case this would suggest that ${\rm N}_{\rm Ca}/{\rm N}_{\rm
Na}\sim0.06$--0.12, i.e.\ a depletion of calcium by a factor $\sim10$. The
ratios observed here suggest clouds of $T_{\rm e}\,\lsim\,7000$ K and
(probably within an order of magnitude) $n_{\rm e}\sim1$ cm$^{-1}$, also on
the basis of calculations by Bertin et al.\ (1993), which is typical of the
warm ionized medium.

Wood \& Bates (1993, 1994) and Bates et al.\ (1992) have probed the ISM in
front of, and surrounding, $\omega$\,Cen with a few dozen stellar probes at
distances from $\sim100$ pc to $>0.5$ kpc. They identified four main kinematic
components which are also spatially distinct: (i) absorption around zero
km~s$^{-1}$ arising from hot gas in the Local Bubble; (ii) absorption around
$-10$ km~s$^{-1}$ arising from cooler material between the Local Bubble and
the expanding Scorpius-Centaurus shell; (iii) strong absorption around $-15$
km~s$^{-1}$, associated with IR emission seen in IRAS maps, in a vertical
extension of the Carina-Sagittarius spiral arm at $\sim1$--2 kpc distance from
the Sun, and (iv) weaker and patchier absorption at more negative velocities,
typically $-40$ km~s$^{-1}$, arising from extra-planar gas beyond the
Carina-Sagittarius arm, at $\gsim\,300$ pc above the Disc. If most of the
absorption seen in the Ca\,{\sc ii} K and Na\,{\sc i} D$_2$ lines is around
1--2 kpc distance then we probe scales of $\sim1$--20 pc.

The mean velocity of the Ca\,{\sc ii} K line is $\langle v_{\rm LSR,
Ca\,II\,K}\rangle=-27$ km~s$^{-1}$. This suggests that the Ca\,{\sc ii} K line
originates predominantly in the extra-planar gas at a few kpc distance, either
because it is warmer there or because dust depletion diminishes the Ca\,{\sc
ii} K line closer to us. The mean velocity of the Na\,{\sc i} D$_2$ line is
$\langle v_{\rm LSR, Na\,I\,D2}\rangle=-14$ km~s$^{-1}$. The Na\,{\sc i} D$_2$
line thus appears to arise predominantly in the puffed-up spiral arm at 1--2
kpc distance. This picture is confirmed by the fact that the more negative
velocity seen in the Ca\,{\sc ii} K line compared to the Na\,{\sc i} D$_2$
line occurs where the Ca\,{\sc ii} K line is broader, the ${\rm W}_{\rm
Ca\,II\,K}/{\rm W}_{\rm Na\,I\,D2}$ ratio is higher and the reddening is
lower.

The kinematic lag observed in the extra-planar gas may not just be due to the
projection of Galactic rotation. Gas infalling from the Halo will lag behind
the Galactic rotation (Albert et al.\ 1993; Shull et al.\ 2009). Crawford et
al.\ (2002) performed a detailed study in the Ca\,{\sc ii} K and Na\,{\sc i}
D$_2$ lines, but at much higher spectral resolution, of similar warm gas in
the lower Halo seen through the Local Interstellar Chimney. They found
evidence of cooling gas precipitating onto the Disc. Alternatively, drag
experienced in the Disc--Halo interaction z\^one where the rotating gaseous
Disc meets the hot Halo gas might slow down rotating extra-planar gas. Such
interaction might also give rise to instabilities in the Disc, resulting in a
spiral density wave, much akin to the standing sound waves generated when
rubbing a metal pipe.

To investigate the possible presence of a velocity lag, we computed the line
profile analytically, from:
\begin{equation}
I(v) = e^{-\tau(v_{\rm LSR})},
\end{equation}
where the optical depth is described as a function of distance to the Galactic
Centre, $r$, and the Galactic plane, $z$, in a double exponential form:
\begin{equation}
\tau(v_{\rm LSR}) = \tau_0\ e^{-r/h_r} e^{-z/h_z} \left|\frac{{\rm d}d(v_{\rm
LSR})}{{\rm d}v_{\rm LSR}}\right|,
\end{equation}
where $h_r$ and $h_z$ are the Disc scalelength and scaleheight, respectively,
and $d$ is the distance from the Sun.

The analytical expression for $d(v_{\rm LSR})$ assuming a flat rotation curve,
$v(r)=v_\odot$, and some of the derived quantities, are discussed in Appendix
A. There, we also describe the more ungainly but still analytical result for
the case where we introduce a velocity lag as a function of $z$ (``drag''):
\begin{equation}
v(z) = v_\odot \left(1-\frac{d}{X_z}\sin(b)\right),
\end{equation}
where $X_z$ sets the rate of linear decline in the velocity. We assumed a
distance to the Galactic Centre of $r_\odot=8$ kpc, $v_\odot=250$ km~s$^{-1}$
(c.f.\ Shattow \& Loeb 2009), and $h_r=3$ kpc. The results are visualized in
Fig.\ 15, compared to the observed Ca\,{\sc ii} K line.

%
\begin{figure}
\centerline{\psfig{figure=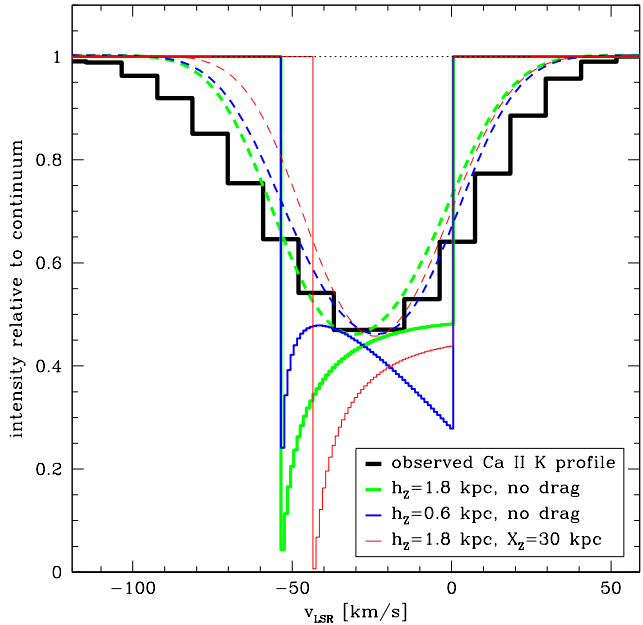,width=84mm}}
\caption[]{Analytical line profiles for different scaleheights, and for the
case of drag exerted by the hot Halo gas. They all peak sharply, either at
$v_{\rm LSR}=-53$ or at $-42$ km~s$^{-1}$. The profiles are also convolved
with an instrumental profile appropriate for our spectra (dashed), and then
peak around $v_{\rm LSR}\sim-30$ km~s$^{-1}$, to be compared with the observed
Ca\,{\sc ii} K profile (coarse histogram).}
\end{figure}

The model with $h_z=1.8$ kpc (c.f.\ Gaensler et al.\ (2008), for the warm
ionized medium) peaks at too negative a velocity. The peak shifts to a more
positive velocity if the scaleheight is lowered, or if some drag is
introduced. The negative-velocity wing of the line profile shifts more in the
latter case. Our average spectrum of the Ca\,{\sc ii} K line is marginally
more consistent with a lower scaleheight without the need to invoke drag. On
the other hand, the most negative component in the higher resolution spectra
presented in Bates et al.\ (1992) peaks at $v_{\rm LSR}=-42$ km~s$^{-1}$,
which is exactly where a model peaks that has a moderate amount of drag
($X_z\sim30$ kpc). In that model, this velocity extremum is reached at $d=4.8$
kpc, i.e.\ just in front of $\omega$\,Cen (assumed to be at $d=5$ kpc).

\subsection{Extra-planar small-scale ISM structure}

As concluded in the previous section, our view towards $\omega$\,Cen seen
through the eyes of the Ca\,{\sc ii} and Na\,{\sc i} lines and $\lambda$5780
and 5797 DIBs mostly shows the spiral arm--Halo interface and the extra-planar
gas beyond it. Although we see neutral atomic and molecular gas, as well as
dust depletion, that portion of the Disc--Halo interaction region has a
prominent component of warm ionized gas, with a scaleheight of $\sim1.8$ kpc
and a filling factor of $\sim30$\% (Gaensler et al.\ 2008) --- note that
$\omega$\,Cen is 1.3 kpc above the Galactic plane. The hot corona permeating
the Halo, making up most of the remaining 70\% in volume, is not probed here.

Even though the ISM is a dynamic environment in which pressure equilibrium is
unlikely to be strictly valid, it is largely the case that the volume filling
factor of an ISM phase matches its temperature, following the equation of
state $P\propto\rho T/\mu$ (with $\mu$ the mean molecular weight). Cold media
are found in the form of compact structures, with a high density (exacerbated
by the lower temperature). But as the cross-section of such structures
decreases, it is harder to sample the colder medium by means of discrete
sightlines (as opposed to full imaging). Furthermore, superposition of
separate cold structures is likely if the column length is substantial.
Variations in column density across the sky should be reflected in variations
along the sightline.

We have performed a simplistic analytical modeling attempt to assess the
small-scale structure of the extra-planar neutral and low-ionized medium. This
assumes a size distribution of spherical clouds of radii $r$, between $r_0$
and $R_0$:
\begin{equation}
N(r) = N_0 \left(\frac{r}{r_0}\right)^{-\xi},
\end{equation}
with the specific density postulated by
\begin{equation}
\rho(r) = \rho_0 \left(\frac{r}{r_0}\right)^{-\chi}.
\end{equation}
The normalisation of the size distribution is obtained from
\begin{equation}
f = \frac{1}{V} \int_{r_0}^{R_0} {\rm d}V(r),
\end{equation}
where $f$ is the volume filling factor for a rectangular volume of space, $V$.
In Appendix B we describe the derivation from this of the expected relative
fluctuations in column density, $\Pi$. For reasonable assumptions of the
parameters upon which this depends (see Appendix B), one would expect $\langle
\sigma_\Pi / \Pi \rangle\sim0.1$--0.2, and this is entirely consistent with
the fluctuations we see in our maps. This once more confirms that we detect
pc-scale structure in the neutral and low-ionized extra-planar gas.

We can not ascertain whether the extra-planar small-scale structure originates
in the Galactic Disc and simply persists as gas migrates to higher elevations,
or whether it is generated {\it in situ}, as a result of thermal instabilities
in cooling gas. The latter may be associated with gas raining down from
outflows generating within the ``Galactic Fountain'' paradigm (Shapiro \&
Field 1976; Spitzer 1990) --- c.f.\ Crawford et al.\ (2002) for recent
evidence of this scenario. Alternatively, it may drizzle down from much
further afield (e.g., Stanimirovi\'c et al.\ 2006; Bland-Hawthorn et al.\
2007).

\section{Conclusions}

Medium-resolution spectra of 452 blue-Horizontal Branch stars in the globular
cluster $\omega$\,Centauri were obtained, with the aim to determine the
spatial and thermodynamical structure of the intervening diffuse interstellar
medium. The study presented here is based on the Ca\,{\sc ii} K and Na\,{\sc
i} D$_2$ lines and the diffuse interstellar bands at 5780 and 5797 \AA, which
probe mainly the warm neutral and weakly-ionized medium. Maps were created
spanning $0.7^\circ$, showing fluctuations in all these tracers, up to a
factor two in the Ca and Na lines, on all scales down to a few arcminutes.
These fluctuations were reproduced in a statistical sense adopting a simple
analytical model for a population of clouds with power-law distributions over
size and density. Correlations with reddening confirmed that calcium depletion
onto dust grains occurs in neutral clouds. The location of the Ca K and Na D
dominant line-forming regions was deduced, on the basis also of kinematical
evidence (line centroids and line widths) in agreement with literature
high-resolution studies of a small number of sightlines.

The most pertinent conclusions of our investigation can be summarised as
follows:
\begin{itemize}
\item[$\bullet$]{In the direction towards $\omega$\,Centauri,
$(l,b)=(309^\circ,15^\circ)$, the Ca K and Na D lines and the 5780 and 5797
\AA\ DIBs trace predominantly the neutral and weakly-ionized medium in the
extra-planar Disc--Halo interface. The neutral medium is relatively more
prominent in the front of this extended region, $\sim400$ pc above the nearby
(1--2 kpc) Carina-Sagittarius spiral arm, whilst the ionized medium dominates
behind it, $\sim0.5$--1 kpc above the Galactic plane.}
\item[$\bullet$]{The extra-planar gas is dominated by $\sigma$-type clouds, in
which the macromolecules held responsible for the DIBs are electrically
charged. The 5780 \AA\ DIB is identified with such carrier; it also appears
stronger in the {\it envelopes} of neutral clouds, i.e.\ in the contact
interfaces with warm gas, whilst the 5797 \AA\ DIB is stronger {\it within}
neutral clouds. This is consistent with the 5797 \AA\ DIB being associated,
either with a neutral carrier molecule or with a singly-charged cation at a
lower photo-ionization threshold of the parent neutral molecule.}
\item[$\bullet$]{Small-scale structure is detected in both the neutral and
weakly-ionized medium, down to scales of $\sim1$ pc. This implies that
small-scale structure persists --- or is generated --- well above the Galactic
plane, several hundreds of pc (or more) away from the nearest sites of
supernovae.}
\end{itemize}

\section*{Acknowledgments}

Jacco wishes to thank Sne\v{z}ana Stanimirovi\'c for inspiration that started
this project. We are grateful for a service time award for AAOmega on the AAT.
We would also like to thank the referee, Ian Crawford, for his constructive
and valuable remarks. KTS thanks EPSRC for a studentship.


\appendix

\section{Computation of line profiles with and without drag}

To compute a line profile (Eq.\ (2)), we need to assess the optical depth in a
velocity interval. This is related to the column density, and thus obtained
from
\begin{equation}
\tau = \tau_0\ n(r,z) \left|\frac{{\rm d}d(v_{\rm LSR})}{{\rm d}v_{\rm
LSR}}\right|.
\end{equation}
In our application, the precise value of $\tau_0$ is not important so we scale
it to match the observed line profile. The density, $n(r,z)$, as a function of
distance to the Galactic Centre, $r$, and Galactic plane, $z$, is described in
our model by a double exponential (c.f.\ Eq.\ (3)). The main mathematical
exercise is to prescribe $r$, $z$, and distance $d$ as a function of $v_{\rm
LSR}$.

Trigonometry relates $r$ and $d$:
\begin{equation}
r^2 = r_\odot^2 + d^2\cos^2(b) -2r_\odot d\cos(l)\cos(b),
\end{equation}
and $z$ and $d$:
\begin{equation}
z = d\sin(b),
\end{equation}
where $l$ and $b$ are the Galactic coordinates of the sightline. Further
simple trigonometry yields a prescription for the radial velocity difference
as a function of $r$ (and thus of $d$):
\begin{equation}
v_{\rm LSR} = \left(v\frac{r_\odot}{r}-v_\odot\right)\sin(l)\cos(b),
\end{equation}
where $v$ and $v_\odot$ are the Galactic rotation velocities of the gas cloud
and the Sun, respectively. For a flat rotation curve, $v(r)=v_\odot$, the
distance at which the velocity difference reaches an extremum is given by
\begin{equation}
d_{\rm x} = r_\odot\frac{\cos(l)}{\cos(b)}.
\end{equation}
For the sightline towards $\omega$\,Cen, $(l,b)=(309^\circ,15^\circ)$, and
adopting $r_\odot=8$ kpc and $v_\odot=250$ km~s$^{-1}$, one obtains $d_{\rm
x}=5.2$ kpc and a corresponding $v_{\rm LSR, x}=-54$ km~s$^{-1}$.

To obtain a prescription for $d(v_{\rm LSR})$, we need to invert Eq.\ (A4)
(incorporating Eq.\ (A2)). For the case of a flat rotation curve this is
straightforward:
\begin{equation}
d(v_{\rm LSR}) =
\left(\cos(l)\pm(a^2-\sin^2(l))^{0.5}\right)\frac{r_\odot}{\cos(b)},
\end{equation}
where we introduce for the sake of brevity:
\begin{equation}
a = \left(\frac{v_{\rm LSR}}{v_\odot\sin(l)\cos(b)}+1\right)^{-1}.
\end{equation}
The plus/minus sign in Eq.\ (A6) arises from the fact that beyond $d_{\rm x}$
the velocity evolves back towards zero, resulting in two solutions for the
distance at any given velocity. In our situation, this is avoided by us only
tracing gas up to the distance of $\omega$\,Cen, which is just before the
turning point $d_{\rm x}$. Hence, in our case the minus sign applies to Eq.\
(A6). The derivative is also readily derived, irrespectively of the sign:
\begin{equation}
\left|\frac{{\rm d}d(v_{\rm LSR})}{{\rm d}v_{\rm LSR}}\right| =
\left|\frac{a^3r_\odot}{v_\odot\sin(l)\cos^2(b) (a^2-\sin^2(l))^{0.5}}\right|.
\end{equation}

The situation becomes more laborious if we replace the flat rotation curve by
the modification following Eq.\ (4), to allow for some kind of drag between
the Disc rotation and the Halo. However, the choice of this form still allows
an analytical solution to be obtained, viz.\ for the distance:
\begin{equation}
d(v_{\rm LSR}= \left(p_3\pm(p_3^2-p_1p_2)^{0.5}\right)\frac{r_0}{p_2\cos(b)},
\end{equation}
and for the derivative:
\begin{equation}
\left|\frac{{\rm d}d(v_{\rm LSR})}{{\rm d}v_{\rm LSR}}\right| =
\left|\frac{r_\odot}{v_\odot\sin(l)\cos^2(b)}(q_1+q_2)\frac{a^3}{p_2}\right|,
\end{equation}
where we have made use of the following abbreviations:
\begin{equation}
k = \frac{r_\odot}{X_z}\tan(b);
\end{equation}
\begin{equation}
p_1 = 1-a^2;
\end{equation}
\begin{equation}
p_2 = 1-k^2a^2;
\end{equation}
\begin{equation}
p_3 = \cos(l)-ka^2;
\end{equation}
\begin{equation}
q_1 = -2\frac{k^2}{p_2}\left(p_3\pm(p_3^2-p_1p_2)^{0.5}\right);
\end{equation}
\begin{equation}
q_2 = 2k\pm(2kp_3-k^2p_1-p_2)(p_3^2-p_1p_2)^{-0.5}.
\end{equation}
Now, the turning point will have shifted, to
\begin{equation}
d_{\rm x} = \left(\frac{\cos(l)-k}{1-k\cos(l)}\right)\frac{r_\odot}{\cos(b)}.
\end{equation}
For moderate drag, with $X_z=30$ kpc, the turning point is located at $d_{\rm
x}=4.8$ kpc, and $v_{\rm LSR, x}=-43$ km~s$^{-1}$. Hence we must add the
contributions from both solutions: the one up to $d=d_{\rm x}$, with the minus
sign in Eqs.\ (A9), (A15) and (A16), and the one from $d_{\rm x}$ until 5 kpc,
with the plus sign.

\section{Derivation of expected column density fluctuations}

The volume contributed by clouds (following Eqs.\ (5)--(7)) of a size between
$r$ and $r+{\rm d}r$ is ${\rm d}V(r) = \frac{4}{3}\pi r^3 {\rm d}N(r)$.
Assuming the volume $V$ has a width $R$ and depth $L$, we obtain
\begin{equation}
N_0 = \frac{3}{4\pi} f \left(\frac{LR^2}{r_0^\xi}\right) A,
\end{equation}
where
\begin{equation}
A = \left\{\begin{array}{ll}
(3-\xi)\left(R_0^{3-\xi}-r_0^{3-\xi}\right)^{-1} & {\rm for}\ \xi\neq 3,\\
\left(\ln(R_0/r_0)\right)^{-1}           & {\rm for}\ \xi=3.
\end{array} \right.
\end{equation}

The expectation value for the column density, $\langle \Pi \rangle$, that we
expect to measure along a given sightline is obtained by integrating the
column density ${\rm d} \Pi$ resulting from clouds of size $r$, which is given
by
\begin{equation}
{\rm d} \Pi = 4 r \rho(r) (r/R)^2\ {\rm d} N(r).
\end{equation}
Here, a factor $\pi (r/R)^2$ was introduced to reflect the cloud's relative
cross section, and a factor $4r/\pi$ to reflect the mean path through a
spherical cloud with radius $r$. Integrating over all cloud sizes, we thus
obtain:
\begin{equation}
\langle \Pi \rangle = 4 \rho_0 N_0 \left(\frac{r_0^{\xi+\chi}}{R^2}\right)
\int_{r_0}^{R_0} r^{2-\xi-\chi}\ {\rm d} r.
\end{equation}

Fluctuations in the number of clouds of a size between $r$ and $r+{\rm d} r$
along a given sightline,
\begin{equation}
\sigma_{{\rm d} N}^2 = \pi \left(\frac{r}{R}\right)^2 {\rm d} N(r),
\end{equation}
cause fluctuations in the column densities amongst different sightlines,
$\langle \sigma_\Pi^2 \rangle$. The expectation value for the variance in
column density is obtained by adding in quadrature the contributions to the
variance by clouds of different sizes, viz.\ $((4 r/\pi) \rho(r) \sigma_{{\rm
d} N})^2$:
\begin{equation}
\langle \sigma_\Pi^2 \rangle = \frac{16}{\pi} \rho_0^2 N_0
\left(\frac{r_0^{\xi+2\chi}}{R^2}\right) \int_{r_0}^\delta r^{3-\xi-2\chi}\
{\rm d} r,
\end{equation}
where $\delta$ is the separation between sightlines. The reason for not
integrating up to $r\geq R_0$ is that sightlines behave covariantly on a scale
that is smaller than the cloud size, hence the contribution of clouds to the
variance diminishes if they are larger than the typical sightline separation.
We thus obtain for the expected relative fluctuations:
\begin{equation}
\langle \sigma_\Pi / \Pi \rangle = \left(\frac{3}{4}\xi f L A\right)^{-1/2} B,
\end{equation}
where
\begin{equation}
B = \left\{\begin{array}{ll}
\left(\frac{\delta^\epsilon-r_0^\epsilon}{\epsilon}\right)^{1/2}
\frac{\eta}{R_0^\eta-r_0^\eta} &
{\rm for}\ \epsilon\neq 0\ \wedge\ \eta\neq 0, \\
 & \\
\frac{(\ln(\delta/r_0))^{1/2}}{\ln(R_0/r_0)} &
{\rm for}\ \epsilon=0\ \wedge\ \eta=0, \\
 & \\
\left(\frac{\delta^\epsilon-r_0^\epsilon}{\epsilon}\right)^{1/2}
\frac{1}{\ln(R_0/r_0)} &
{\rm for}\ \epsilon\neq 0\ \wedge\ \eta=0, \\
 & \\
(\ln(\delta/r_0))^{1/2} \frac{\eta}{R_0^\eta-r_0^\eta} &
{\rm for}\ \epsilon=0\ \wedge\ \eta\neq 0.
\end{array} \right.
\end{equation}
where $\epsilon = 4-\xi-2\chi$ and $\eta = 3-\xi-\chi$.

As an example, for $\xi=2$ and $\chi=1$ (c.f.\ Kim et al.\ 2007), i.e.\
$\epsilon=\eta=0$, and reasonable values for $R_0\sim10$ pc, $r_0\sim1$ AU,
$\delta\sim6$ pc ($10^\prime$ at 2 kpc)\footnote{For $n$ sightlines
distributed randomly over an area of $R^2$, the average separation is $\langle
\delta \rangle \simeq R/\sqrt{n}$.}, $L\sim2$ kpc, and $f\sim0.3$, we obtain
$\langle \sigma_\Pi / \Pi \rangle\simeq0.084$. It largely depends on
$(fL/R_0)^{-1/2}$ (as long as $R_0\gg r_0$), so for a larger cut-off in cloud
size and/or a shorter characteristic column the value for $\langle \sigma_\Pi
/ \Pi \rangle$ increases. As $R_0\leq R$, with $R\sim25$ pc ($0.7^\circ$ at 2
kpc), and $L$ is unlikely to be much less than a kpc, $\langle \sigma_\Pi /
\Pi \rangle$ could possibly reach $\sim0.2$. It also increases if the filling
factor is smaller, so if the contribution of colder neutral medium is
relatively more important than the warm ionized medium then $f<0.3$. This is
likely to be the case, as we saw in the previous sections that we probe a
mixture of neutral and low-ionized material (c.f.\ Welsh, Wheatley \&
Lallement 2009, who probe similar conditions).

\label{lastpage}

\end{document}